\pdfoutput=1

\documentclass[11pt,twoside,a4paper,cmspaper,final,collab]{cms-tdr}
\def\svnVersion{325dab2-D}\def\svnDate{2020/05/05}\def\cmsCernNoTag{CERN-EP-2019-179}\def\cmsCernDate{\today}\def\cmsMessage{"Published in the Journal of Instrumentation as \href{http://dx.doi.org/10.1088/1748-0221/15/05/P05002}{\doi{10.1088/1748-0221/15/05/P05002}.}"}

\begin{document}\cmsNoteHeader{PRF-18-001}

\hyphenation{had-ron-i-za-tion}
\hyphenation{cal-or-i-me-ter}
\hyphenation{de-vices}
\RCS$HeadURL: svn+ssh://svn.cern.ch/reps/tdr2/papers/PRF-18-001/trunk/PRF-18-001.tex $
\RCS$Id: PRF-18-001.tex 493532 2019-04-06 20:54:22Z banerjee $

\providecommand{\cmsLeft}{left\xspace}
\providecommand{\cmsRight}{right\xspace}
\newcommand{\pu}{pileup}
\newcommand{\zee}{\ensuremath{\PZ \to \Pe\Pe}\xspace}
\newcommand{\zmm}{\ensuremath{\PZ \to \MM}\xspace}
\newcommand{\etot}{\ensuremath{E_\text{tot}}\xspace}
\newcommand{\ieta}{\ensuremath{i\eta}\xspace}
\newcommand{\iphi}{\ensuremath{i\phi}\xspace}

\cmsNoteHeader{PRF-18-001}

\title{Calibration of the CMS hadron calorimeters using proton-proton collision data at $\sqrt{s} = 13\TeV$}

\date{\today}

\abstract{
Methods are presented for calibrating the hadron calorimeter system of the CMS detector at the LHC. The hadron calorimeters of the CMS experiment are sampling calorimeters of brass and scintillator, and are in the form of one central detector and two endcaps. These calorimeters cover pseudorapidities $\abs{\eta} < 3$ and are positioned inside the solenoidal magnet. An outer calorimeter, outside the magnet coil, covers $\abs{\eta} < 1.26$, and a steel and quartz-fiber Cherenkov forward calorimeter extends the coverage to $\abs{\eta} < 5.19$. The initial calibration of the calorimeters was based on results from test beams, augmented with the use of radioactive sources and lasers. The calibration was improved substantially using proton-proton collision data collected at $\sqrt{s} = 7$, 8, and 13\TeV, as well as cosmic ray muon data collected during the periods when the LHC beams were not present. The present calibration is performed using the 13\TeV data collected during 2016 corresponding to an integrated luminosity of 35.9\fbinv. The intercalibration of channels exploits the approximate uniformity of energy collection over the azimuthal angle. The absolute energy scale of the central and endcap calorimeters is set using isolated charged hadrons. The energy scale for the electromagnetic portion of the forward calorimeters is set using \zee data. The energy scale of the outer calorimeters has been determined with test beam data and is confirmed through data with high transverse momentum jets. In this paper, we present the details of the calibration methods and accuracy.
}

\hypersetup{
pdfauthor={CMS Collaboration},
pdftitle={Calibration of the CMS hadron calorimeters using proton-proton collision data at sqrt(s) = 13 TeV},
pdfsubject={CMS},
pdfkeywords={CMS, physics, detector, hadron calorimeter}
}

\maketitle

\section{Introduction}\label{sec:intro}

Most precision studies of the standard model (SM) utilizing proton-proton
({$\Pp\Pp$}) collisions depend on reliable, precise measurements of jets and
missing transverse momentum.
The uncertainties in these are often related to the energy scale and
the resolution of the measurement of hadron energy.
For example, a precision measurement of the mass of the top quark requires a
detailed understanding of the energy scale and resolution of jets; in a recent
measurement of the top quark mass by CMS~\cite{PhysRevD93072004,Sirunyan2019EPJ313},
the single largest source of experimental uncertainty in two of the
channels used was the  jet energy scale.
Searches for physics beyond the SM (\eg, searches for dark matter based on
measurements of jets produced through initial-state radiation~\cite{EPJC752352015})
often rely on measurements of hadron properties.

The methods used to calibrate hadron calorimeters have been
studied and improved by several experiments.
Previous publications from the CMS and ATLAS Collaborations on calorimeter
calibration are reported in~Refs.~\cite{1748-0221-11-10-T10005,Aaboud2017,TB96,HCalOuter,TB06,COJOCARU2004481}.
The hadron calorimeter (HCAL)~\cite{HcalTDR,CMS} is composed of four major subdetectors: the hadron barrel (HB)~\cite{HcalBarrel}, the hadron endcap (HE)~\cite{HcalTDR}, the hadron forward (HF)~\cite{HcalForward}, and the hadron outer (HO) calorimeters~\cite{HCalOuter} (as shown in Fig.~\ref{fig:hcal}).
The CMS Collaboration has developed several techniques for calibrating its HCAL.
The initial calibration makes use of results from several test beam exposures and from signals injected by dedicated calibration systems based on lasers and radioactive sources.
Details of the test beam analyses and the calibration using radioactive sources can be found in Refs.~\cite{TB96,TB06,WIRE}.
The calibration of the detector is improved through the analysis of data from
cosmic ray muons, taken when the CERN LHC was not operating, and data containing
energy deposits from secondary particles produced during LHC beam
tuning~\cite{HcalCalib1} and traversing the detector  longitudinally.

The final calibration, described in this paper, uses information from
collision data to further improve the precision of the calibration, and
to establish a hadronic energy scale, which is stable over the course of the data taking.
Because of the complex structure of the HCAL, its large angular coverage, nonuniformities in the amount of material in front of the calorimeters, and the limited acceptance of the CMS tracking system, these goals can only be achieved through the use of several techniques and data samples.
In addition, the calibration needs to take into account
nonlinearities in the HCAL energy response~\cite{TB06}.
The calibration methods reported in this paper were first used with data
collected  during 2010--2015 at $\sqrt{s} = 7$, 8, and 13\TeV.
Here, we report the calibration performed using data collected during 2016
at $\sqrt{s} = 13\TeV$, corresponding to an integrated luminosity up to
35.9\fbinv.

Detector calibration is needed to reduce the uncertainty in the energy measurement to less than 3\%. Sources of changes in response larger than this target goal include aging of the photocathodes of the photodetectors, scintillator and fiber optic aging, and changes to the configuration of the hardware during shutdown, {\it e.g.}, the readout of the HF calorimeter was split into two separate optical paths during a shutdown at the end of 2015.

Because of the nonlinear energy response of the HCAL, it is not possible to set an absolute energy scale that is valid for all incident hadron momenta.
Although the response is closer to being linear at high momenta,
there are relatively few events available. Therefore, the calibration is
performed with a sample of isolated hadrons of moderate momentum.
The calibration of the HB and HE described in this paper yields a unit value
for the relative energy scale factor of isolated charged pions with momenta
of 50\GeV, which do not interact hadronically in the CMS electromagnetic
calorimeter (ECAL).
The energy of a reconstructed particle~\cite{CMS-PRF-14-001} is corrected for the calorimeter nonlinearity
using a parametrization of the response as a function of
transverse momentum (\pt), pseudorapidity ($\eta$),
and the fractions of the particle energy deposited in the different
subdetectors, as determined from a {\GEANTfour}-based~\cite{GEANT} CMS detector
simulation.
Residual nonlinearities for jets are removed during the calibration of the jet energy scale~\cite{JESCMS:2017}.
The low-level calibration, reported in this paper, followed by the corrections
and high-level calibration~\cite{CMS-PRF-14-001,JESCMS:2017}, lead to the final
jet energy calibration of the CMS calorimeter system.

Unlike the HB and HE, the HF does not have a strong nonlinear energy response.
The measured e/$\PGp$ response decreases from 1.14 to 1.01 when the energy of
the incident particles increases from 30 to 150\GeV~\cite{HcalForward}.
The energy scale for the HF is set to have a unit value of the relative energy
scale factor for 100\GeV\ electrons and $\PGpm$ mesons using test beam data~\cite{HcalForward}.

The HO is used to measure the energy deposits from high-\pt\ particles whose showers are not fully contained in the ECAL and HB.
The scale for the HO is set to give the best possible energy resolution for 300\GeV $\PGpm$ mesons~\cite{HCalOuter}.

The  calibration includes the following steps:
\begin{itemize}
 \item[(i)] the responses of different channels at the same
       $\eta$ are equalized  in the HB, HE, and HF, exploiting the approximate
       uniformity of collected energies over the azimuthal angle ($\phi$);
 \item[(ii)] for the HB and the part of HE within the
       acceptance of the CMS tracking system \mbox{($\abs{\eta} < 2.5$),}
       isolated charged hadrons are used to equalize the $\eta$ response
       and to set the absolute energy scale;
 \item[(iii)] the calibration for the electromagnetic portion of the HF
       is validated using electrons from decays of \PZ bosons; and
 \item[(iv)]the responses of the HO channels are equalized across $\phi$
       using muons from collision data, and the energy scale factor is
       validated using dijet events.
\end{itemize}

The scale factors obtained from the calibration process are updated if they differ from the current set of scale factors by more than 3\%. During 2016, only two updates were necessary for the final processing of the data.

The paper is organized as follows.
Section~\ref{sec:hcal} provides a brief description of the HCAL.
Section~\ref{sec:recon} describes the reconstruction of HCAL
energies and of the physical objects used in the analysis.
Section~\ref{sec:simulation} gives information about the simulated event
samples used in the design and testing of the calibration methods.
The analysis of the symmetry in the azimuthal angle is in Section~\ref{sec:phisym},
and the calibration using isolated tracks is discussed in Section~\ref{sec:isotk}.
The HF calorimeter calibration using \zee events appears in Section~\ref{sec:hf}, and
the intercalibration of the HO calorimeter and the validation of its energy  scale factors
are described in Section~\ref{sec:ho}.
A brief summary of the results is given in Section~\ref{sec:summary}.

\section{The CMS detector and its hadron calorimeter}\label{sec:hcal}

The central feature of the CMS apparatus is a superconducting solenoid of
6\unit{m} internal diameter, providing a magnetic field of 3.8\unit{T}.
Within the solenoid volume are a silicon pixel and strip tracker, a lead
tungstate crystal ECAL, and the HB and HE calorimeters.
Muons are detected in gas-ionization chambers embedded in the steel flux-return yoke outside the solenoid.
In the barrel section of ECAL, an energy resolution of about 1\% is achieved for unconverted or late-converting photons which have energies in the range of tens of \GeV.
For the remaining barrel photons the resolution is about 1.3\% up to $\abs{\eta} = 1$, deteriorating to about 2.5\% at $\abs{\eta} = 1.4$.
In the endcaps, the resolution for unconverted or late-converting photons is
about 2.5\%, whereas the resolution for the remaining endcap photons is
between 3 and 4\%~\cite{JINST08010}.
For isolated particles of $1 < \pt < 10\GeV$ and $\abs{\eta} < 1.4$, the track
resolutions are typically 1.5\% in $\pt$ and 25--90 (45--150)\mum
in the transverse (longitudinal) impact parameter~\cite{TRK-11-001}.
When measurements from the tracker and the calorimeters are used, the jet energy
resolution amounts typically to 15\% at 10\GeV, 8\% at 100\GeV, and
4\% at 1\TeV, to be compared to about 40, 12, and 5\% obtained when
only the ECAL and the HCAL calorimeters are used~\cite{CMS-PRF-14-001}.

Events of interest are selected using a two-tiered trigger
system~\cite{Khachatryan:2016bia}. The first level (L1), composed of custom
hardware processors, uses information from the calorimeters and muon detectors
to select events at a rate of around 100\unit{kHz} within a time interval of
less than 4\mus. The second level, known as the high-level trigger (HLT),
consists of an array of processors running a version of the full event
reconstruction software optimized for fast processing, and reduces the event
rate to around 1\unit{kHz} before data storage.

A more detailed description of the CMS detector, together with a definition of
the coordinate system used and the relevant kinematic variables, is reported
in Section 5 of Ref.~\cite{CMS}.

\begin{figure}[htbp]
\begin{center}
  \includegraphics[width=.84\textwidth]{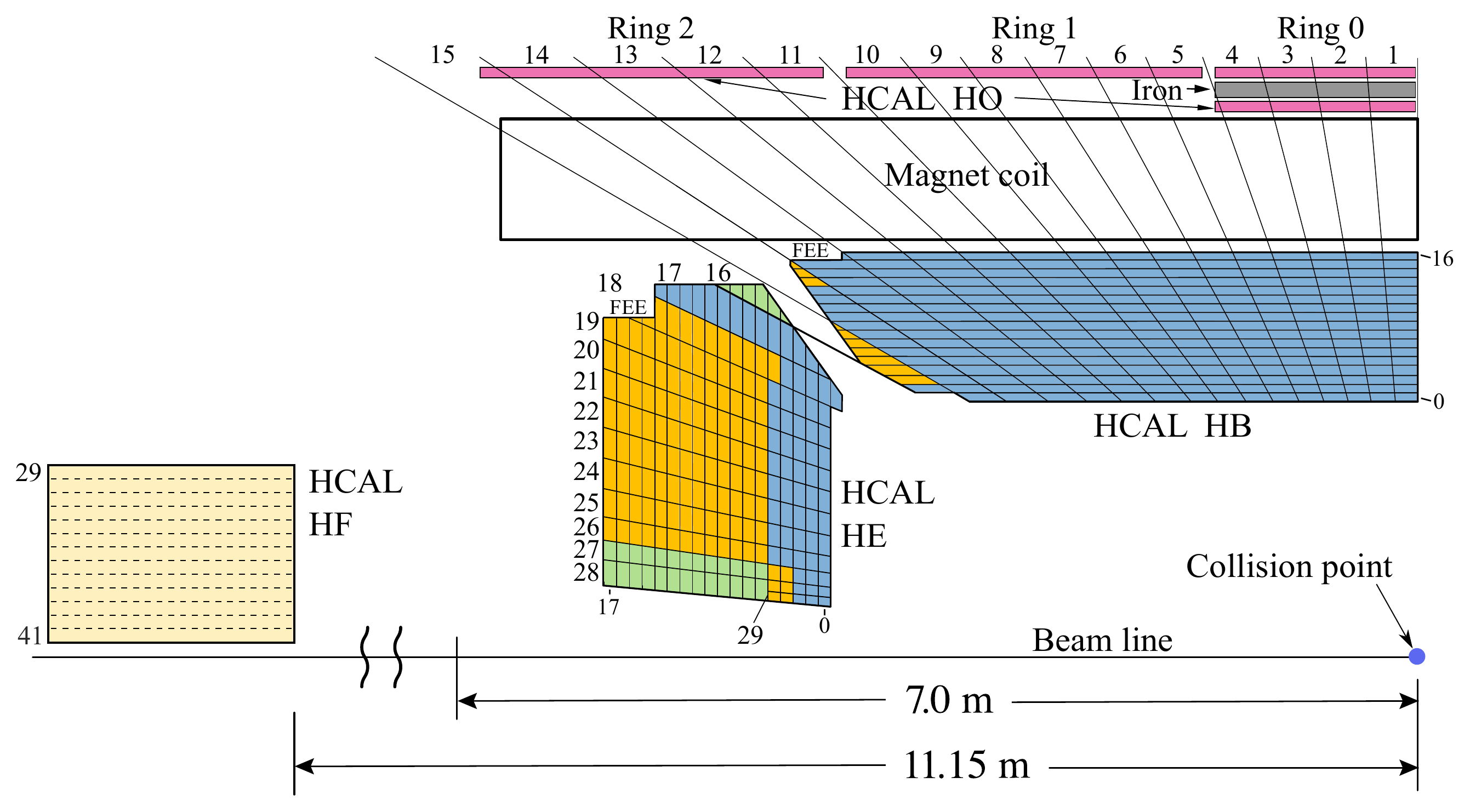}
 \end{center}
 \caption{A schematic view of one quarter of the CMS HCAL during 2016
          LHC operation, showing the positions of its four major components:
          the hadron barrel (HB), the hadron endcap (HE), the hadron outer (HO),
          and the hadron forward (HF) calorimeters. The layers marked in blue
          are grouped together as $\text{depth} = 1$, while the ones in yellow, green,
          and magenta are combined as depths 2, 3, and 4, respectively.}
 \label{fig:hcal}
\end{figure}

Figure~\ref{fig:hcal} shows a schematic view of the layout of the CMS HCAL
during the 2016 LHC operation at $\sqrt{s} = 13\TeV$.
The HB is located between radii of 1775 and 2876.5\mm and covers
$\abs{\eta} < 1.39$.
The HB is divided into two half-barrels in the direction along the
beam ($z$), each assembled from 18 wedges.
Each wedge subtends $20^\circ$ in $\phi$ and extends
to 4330\mm from the CMS detector mid-plane.
A wedge contains absorber plates made of brass (an alloy with 70\% copper and
30\% zinc) that are bolted together.
The inner and outer plates are made out of stainless steel.
There are 17 slots per wedge that house the plastic scintillator tiles.
The inner and outer slots are 14\mm thick while the remaining ones are
9.5\mm thick.
The HB has about $40\,000$ scintillator tiles. In order to limit the number
of individual physical elements, the tiles at the same $\phi$
and depth are grouped into a single scintillator unit, referred to
as a megatile. The megatiles in the first and last layers are of 9\mm
thickness, while the remaining layers have 3.7\mm thick megatiles.
Each megatile covers roughly $5^\circ$ in $\phi$. Of the four
$\phi$ segments within a barrel wedge, the two segments at a larger radius
are staggered with respect to the inner two. There is 61\mm of stainless
steel between layers 0 and 1.
There are 50.5\mm thick brass plates between adjacent layers 1--9, and
the 56.5\mm thick brass plates up to layer 15.
The back plate, which is in front of the last HB calorimeter layer,
is made of 75\mm thick stainless steel.
The megatiles are divided into 16 sections along the $z$~axis, denoted by
$\abs{\ieta} = 1$ to 16, so that each tile corresponds to $\Delta\eta$ of
0.087.
The set of scintillators corresponding to the same value of $\ieta$
and $\iphi$ (denoting the $\phi$ segment) in different layers are grouped
together and referred to as a ``tower''.
All 17 layers are grouped into a single readout channel until $\abs{\ieta} = 14$,
beyond which there are two depth sections, as shown in Fig.~\ref{fig:hcal}.

The HE calorimeter is also made of brass absorber plates with sampling layers of plastic scintillator.
The innermost surface of HE is
located 4006.5\mm from the interaction point and covers
$1.30 < \abs{\eta} < 3.00$. Each endcap has an 18-fold
symmetry in $\phi$ and has 34.5\mm thick sector plates each covering
$20^\circ$ in $\phi$. The sector layers are separated by 9\mm thick brass
spacers to allow space for the scintillator inserts.
Each scintillator insert covers $10^\circ$ in $\phi$.
The top edge of the front part of the endcap module has a slope of $53^\circ$
corresponding to the gap angle between the HB and HE calorimeters. It also has a nose-like
structure, with an additional layer of absorber and scintillator plate
for $\abs{\ieta}$ of 18, to increase the total interaction length for that
tower.
The absorber thickness between successive layers amounts to 78\mm of
brass and scintillator, corresponding to the thickness of two sector plates and one spacer. The
endcap on either side is divided into 14 parts along $\abs{\eta}$, and the 18 layers
are combined into 1, 2, or 3 depth sections, as shown in Fig.~\ref{fig:hcal}.

The HO calorimeter consists of one or two layers of
scintillator outside the magnet coil. The entire assembly is divided
into 5 rings, each having 12 sectors in $\phi$. Six trays of scintillators are
assembled on a honeycomb structure, which is then mounted in each of these
sectors. The central ring (ring 0) has two layers of 10\mm thick
scintillator on either side of a stainless steel block
at radial distances of 3850 and 4097\mm, respectively.
All other rings have a single layer at a radial distance of 4097\mm.
The $\eta$-$\phi$ segmentation of the HO calorimeter matches closely
that of the HB calorimeter.  The HO calorimeter covers $\abs{\eta} < 1.26$.

For the HB and HE calorimeters, clear fibers carry the light to hybrid photodiodes
(HPDs)~\cite{Cushman},
and each HPD signal is digitized in 25\unit{ns} time intervals
by a charge integrator and encoder (QIE)~\cite{qie}.
For the HO calorimeter, light is carried to silicon photomultipliers (SiPMs)
and the SiPM signals are digitized by the QIEs.

The front faces of the HF calorimeters are located 11150\mm away from the
interaction point on either side of the CMS detector and cover
$2.85 < \abs{\eta} < 5.19$.
The detectors covering positive and negative $\eta$ ranges are referred to as
HF$+$ and HF$-$.
The inner and outer radii of the HF calorimeter are 125 and 1570\mm, respectively.
Each HF module is composed of 18 wedges
made of steel with quartz fibers embedded along its length.
The detection technique utilizes emission of Cherenkov light by
secondary charged particles going through the quartz fibers.
Long (1649\mm)
and short (1426\mm) quartz fibers are placed alternately with a separation
of 5\mm. The long fibers reach the front face of the calorimeter, while
the short fibers start 12.5 radiation lengths into the calorimeter.
These fibers are bundled at the back and led to a photodetector,
and are grouped into 13 divisions in $\eta$ on either side of the
CMS detector and 36 divisions in $\phi$, except the two largest $\abs{\eta}$
sections, which contain 18 divisions in $\phi$.
While the HB and HE calorimeters work in conjunction with the ECAL for
particle measurement, the difference in the energy deposits in the long and
short fibers of the HF calorimeter functions as a separator between
electromagnetic and hadronic showers.
The light collected from an HF calorimeter fiber is converted to charge by
a photomultiplier tube and digitized by the QIE.

\section{Event reconstruction}
\label{sec:recon}

In recorded events,
signals from the HB, HE, and HO calorimeters
are stored as 10 consecutive QIE measurements (``time samples'',
where each sample spans 25 \unit{ns}).
To reduce data volume and readout latency, a ``zero suppression''
mechanism is introduced that passes to the data acquisition system only those
channels with at least one set of two consecutive samples above a threshold.
The timing of the readout system is adjusted so that
the triggered bunch crossing is in the fifth time sample.
The data recorded from the HF channels are stored as four consecutive time
samples, with the triggered bunch crossing in the third time sample.
Energies are calculated as sums of time samples, after subtraction of the
electronic pedestal and application of the scale factor.
For the HB and HE, and for the analysis described in Section ~\ref{sec:moments},
the energy is calculated by applying a scale factor to the sum of energies in
the fifth- and sixth-time samples, with a correction factor to account for
the pulse extending beyond 50\unit{ns}. For the HF, since the Cherenkov pulse
is shorter than 25\unit{ns}, only one time sample is used.

A sophisticated reconstruction method is used for the HB and HE subsystems
because there is a large contribution from pileup, and the HPDs used for their
readout have larger noise than the SiPMs used in the HO. For the HB and HE
the energy deposited in the triggered bunch crossing, after subtracting the
electronic pedestal, is obtained by fitting the time samples with up to three
pulse shape templates shifted in time relative to each other.
The reconstruction algorithm also corrects for the residual contributions
caused by additional interactions in the triggered beam crossing as well as
the preceding and subsequent beam crossings (out-of-time pileup).  This
procedure is employed for the standard reconstruction of HB and HE in the
physics analyses for all the results in this paper, except for the process
described in Section ~\ref{sec:moments}, which uses a simple sum of energies
over the two time slices.

The reduction of the light output of the scintillator caused by radiation
damage ~\cite{1748-0221-11-10-T10004} and the decrease in the quantum
efficiency of the HPD photocathodes due to the ion feedback
damage ~\cite{Dugad} are monitored using a laser calibration system, and
corrections are applied to the reconstructed hit energies based on the
monitored results.

Because the HO subsystem is outside the solenoid, the effect of pileup is relatively small, and a simpler procedure is used.  For HO signal pulses longer than 100\unit{ns}, a correction is applied to account for the signal beyond the eighth time sample.

There are several sources of noise in the hadron calorimeter, such as noise in the readout system, as well as noise from physical sources like particles other than optical photons interacting directly with the readout system~\cite{Collaboration_2010}.

Noise caused by spurious electronic signals is detected and subtracted during the calibration procedure. Although the overall rate is low, this noise would result in large values of mismeasured missing transverse momentum so dedicated filters are used to eliminate either the specific calorimeter channel, or the entire event~\cite{Sirunyan_2019}.

The reconstruction of physics objects produced in {$\Pp\Pp$} collisions
starts with the reconstruction of  particles.
The CMS global event description is based on the particle-flow (PF)
algorithm~\cite{CMS-PRF-14-001}, which reconstructs and identifies each
individual particle in an event with an optimized combination of information
from the various elements of the CMS detector.
In this process, the identification of the particle type (photon, electron, muon, charged or neutral hadron) plays an important role in the determination of the particle direction and energy.
Photons, \eg, coming from \Pgpz\ decays or from electron bremsstrahlung, are identified as ECAL energy clusters not linked to any charged-particle trajectory extrapolated to the ECAL.
Electrons, \eg, coming from photon conversions in the tracker material or from \PB\ hadron semileptonic decays, are identified as primary charged-particle tracks and potentially many ECAL energy clusters corresponding to this track extrapolation to the ECAL and to possible bremsstrahlung photons emitted along the path through the tracker material.
Muons, \eg, from \PB\ hadron semileptonic decays, are identified as tracks in the central tracker consistent with either a track or several hits in the muon system that are associated with very low energy deposits in the calorimeters.
Charged hadrons are identified as charged-particle tracks that are not identified as electrons or muons.
The energy of a charged hadron is determined from the combination of the measurements from the tracker and the calorimeters.
Finally, neutral hadrons are identified as HCAL energy clusters not linked to any charged-hadron trajectory, or as an excess in the combined ECAL and HCAL energy with respect to the expected charged-hadron energy deposit.
The reconstructed particles are referred to as PF candidates.

For each event, hadronic jets (PF jets) are clustered from the PF candidates
 using the infrared and collinear safe anti-\kt
algorithm~\cite{Cacciari:2008gp, Cacciari:2011ma} with a
distance parameter of 0.4. The jet momentum is determined as the
vectorial sum of all particle momenta in the jet, and is found from
simulation to be, on average, within 5 to 10\% of its true
momentum over the entire \pt\ spectrum and detector acceptance.
The PF candidates associated with a jet are referred to as jet constituents.
The PF missing transverse energy is calculated from the jet momenta as an
imbalance of the energy flow in the transverse plane of the detector.
Because PF makes optimal use of the detector information, contributions to the
jet and missing transverse energy resolutions from charged hadrons are
dominated by the precision of the momentum measurement from the tracker.
The ECAL determines the contribution to the energy resolution from photons.
However, for the highest energy charged hadrons, which can be produced in
jets originating from decays of high-mass new particles, for neutral hadrons,
and for all particles at large rapidities, the hadron calorimeter dominates
the momentum measurement.
Accurate measurement of HCAL energies is also important in isolation
calculations, which are used in particle identification.

During the 2016 data taking, the mean number of {$\Pp\Pp$} interactions per bunch crossing was approximately 23.
The reconstructed vertex with the largest value of summed physics-object
$\pt^2$ is identified as the primary $\Pp\Pp$ interaction vertex.
The physics objects in this case are jets, clustered using the jet finding algorithm~\cite{Cacciari:2008gp,Cacciari:2011ma} with the tracks assigned to the vertex as inputs, and the associated missing transverse momentum, taken as the negative vector \pt sum of those jets.
Other reconstructed vertices are referred to as \pu\ vertices.

\section{Simulated event samples}
\label{sec:simulation}

The methodologies used for the calibration are tested
using simulated samples of {$\Pp\Pp$} interactions.
Simulation of SM processes, unless otherwise stated, is performed with
\PYTHIA~8.206~\cite{Sjostrand:2014zea} or \MGvATNLO~2.2.2~\cite{Alwall:2014hca}
event generators at leading order in the strong coupling,
which is set to 0.130 at the \PZ boson mass scale.
The event generators employ the \textsc{NNPDF3.0}~\cite{Ball:2014uwa} parton distribution functions.
Parton shower development and hadronization are
simulated with {\PYTHIA} using the underlying event tune CUETP8M1~\cite{Khachatryan:2015pea}.
Simulated samples consisting of single high-\pt\ particles are also produced.
Samples that do not contain a collision in the nominal bunch crossing are used to
simulate noise (see Section~\ref{sec:moments}).
The detector response is simulated using a detailed
description of the CMS detector implemented with the \GEANTfour
package~\cite{GEANT}. The simulated events are reconstructed with
the same algorithms used for the data.
The simulated samples include pileup with the distribution matching
 that observed in the data.

\section{Calibration using azimuthal symmetry}\label{sec:phisym}

The first step in the calibration of the HB, HE, and HF with collision
data is to equalize the response in $\phi$ for each $\ieta$ ring and
depth section. The procedure
takes advantage of the $\phi$-symmetric particle energies and the
corresponding $\phi$-symmetric collected energy from minimum bias (MB)
events (events selected with triggers designed to collect
inelastic collisions with maximum efficiency while suppressing noncollision
events).
The layouts of the barrel and the endcap detectors have some
$\phi$ dependence because of the absorber structure;
the scintillator layers are staggered, and the
absorber layers are also used as a part of the support structure.
For the forward calorimeter, a radial shift in the beam spot position may also
introduce asymmetry in the $\ieta$ rings close to the beam pipe,
which can change with time.
The relative contributions to the $\phi$ asymmetries from materials,
inhomogeneous magnetic field, beam spot
shift, and miscalibration could, in principle, be understood using simulation.
However, the material description in the simulation
and the modeling of the beam spot position is not exact,
and the difference between the actual detector and its
Monte Carlo description can increase with time because of stresses from the
magnetic field, gravity, \etc

Therefore, the intercalibration is performed by comparing the average collected
energy in a calorimeter channel to the average collected energy in the entire
$\ieta$ ring. Two different calibration procedures are adopted.
\begin{description}
\item[Iterative method:]
A set of multiplicative correction factors (scale factors) for the
uncalibrated energies
are determined iteratively by equalizing
the mean of the measured energies that satisfy both an upper and a lower threshold.
This method works best for energies above 4 GeV.

\item[Method of moments:]
This intercalibration is carried out using MB
events taken without zero suppression
by comparing the first (mean) and second (variance) moments of the energy
distribution in a
calorimeter channel to the mean of the moments of the energy distributions in
the entire $\ieta$ ring.
This method works best for low energies, down to a fraction of a GeV.
\end{description}
By construction, these two methods use events from disjoint data samples and are statistically independent.

The method of moments is performed using events with no zero suppression (NZS).
The measured energies include contributions both from genuine energy deposits and from noise.
The contributions due to noise are estimated from an independent data sample taken when there were
no beam collisions (pedestal data), and subtracted from the measurements
made using the collision data to extract the contribution due to signal.
The iterative method, on the
other hand, makes use of zero-suppressed events and is based on an estimation
of mean energy in an energy interval.
The two methods symmetrize different energy ranges and offer the best performance in the corresponding range.

\subsection{Iterative method}\label{sec:iterative}

The data used in this method are selected from events triggered by
subdetectors other than the HCAL
to avoid trigger bias in the energy measurement;
only events collected with electron, photon, and muon triggers are used
(single-muon triggers with $\pt > 24$\GeV and single-electron triggers with
$\pt > 25$\GeV).
Results obtained from different triggers are compared, and the difference
is used as input to the estimation of the systematic uncertainty.

\begin{figure}[htbp]
\begin{center}
  \includegraphics[width=.49\textwidth]{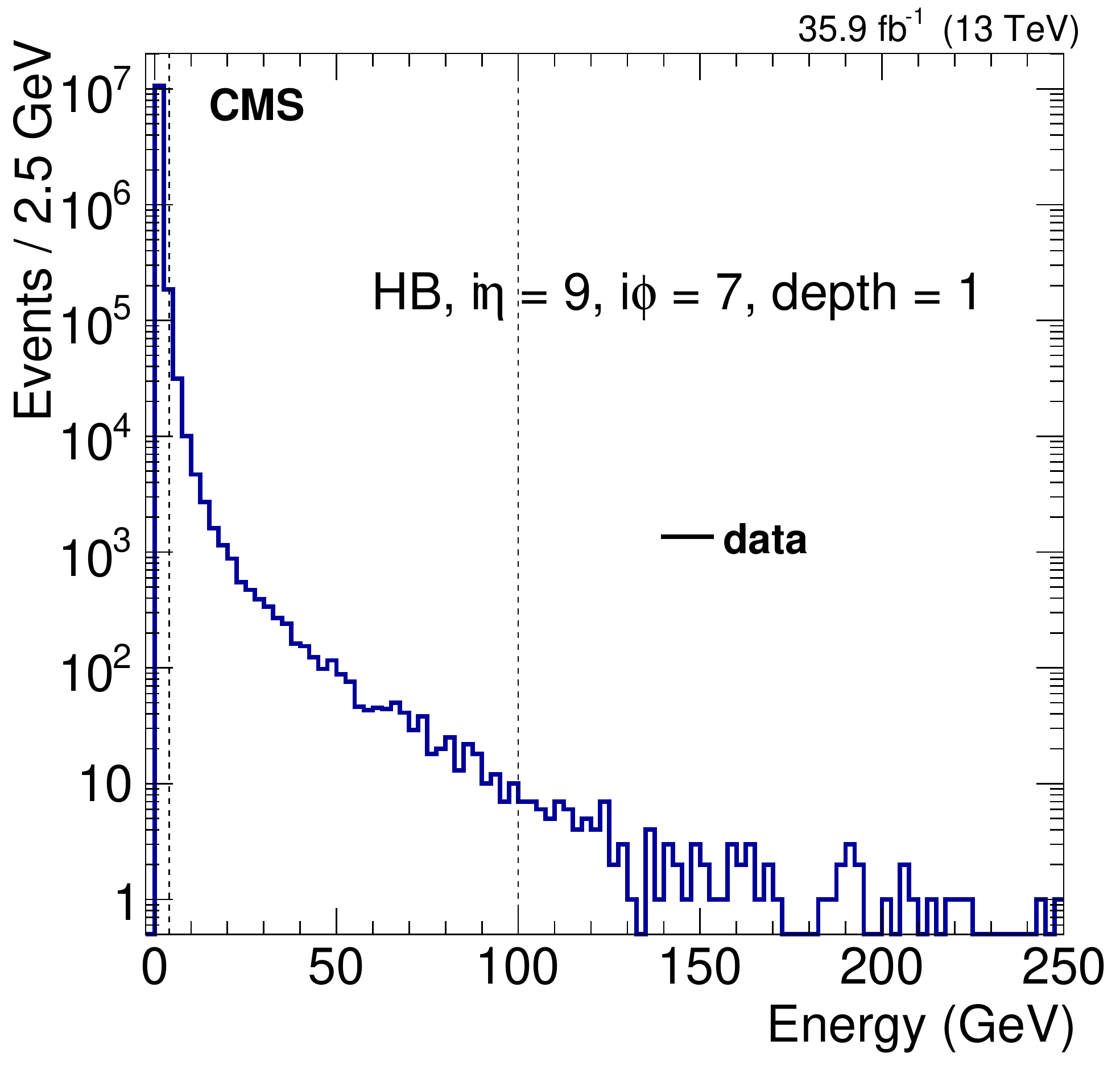}
  \includegraphics[width=.49\textwidth]{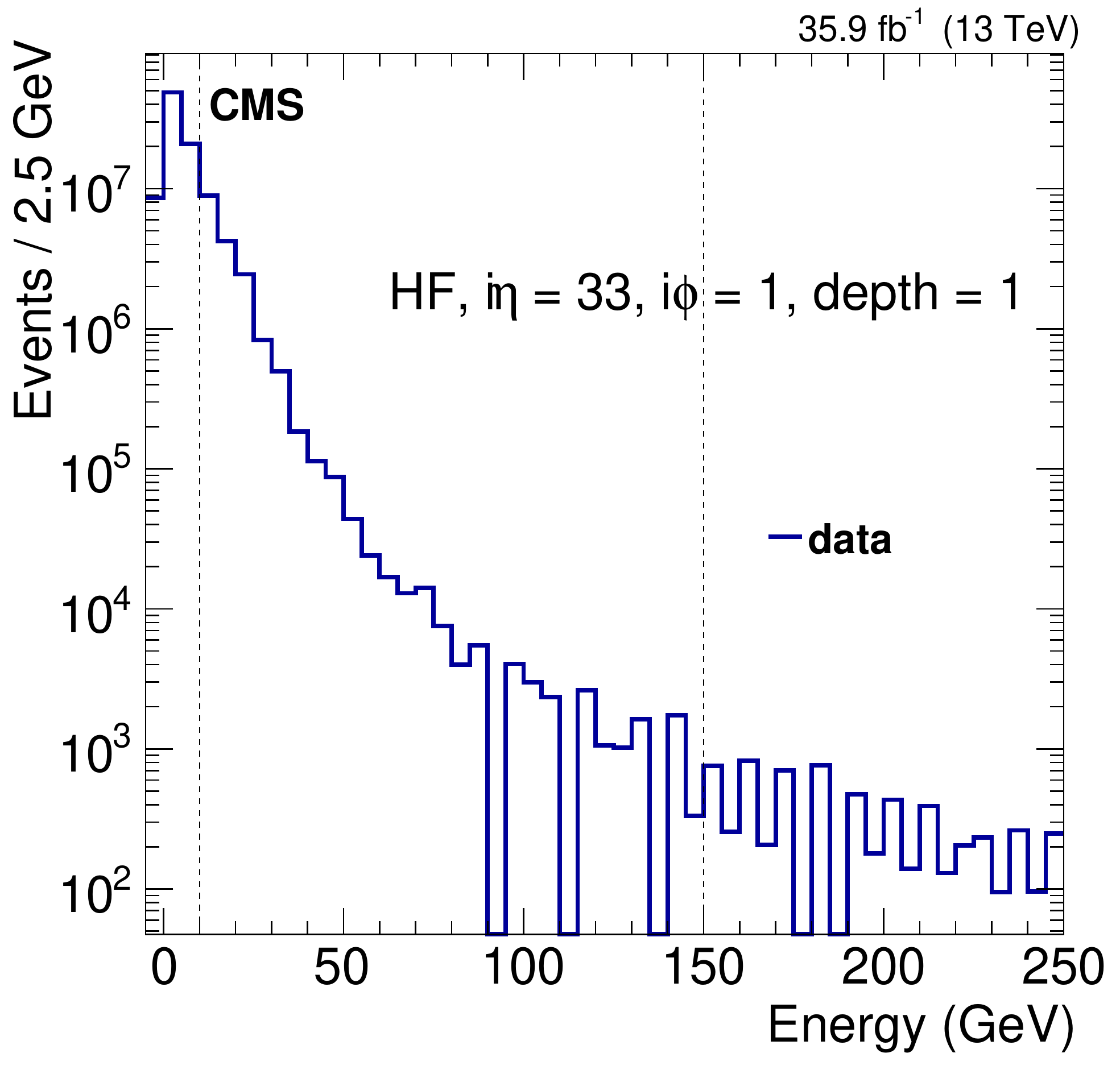}
 \end{center}
 \caption{Energy spectra used as an input to the iterative method of equalizing
          the $\phi$ response of one typical $\ieta$ ring for an HB (\cmsLeft)
          and HF (\cmsRight) calorimeter channel.
          Thresholds on the collected energies used for the energy
          estimation are shown with dashed lines. The legends show the HCAL
          channel index in $\ieta$, $\iphi$, and depth segmentation units.}
 \label{fig:edist}
\end{figure}

This method utilizes energies recorded in HCAL channels that pass
lower and upper threshold requirements.
Figure~\ref{fig:edist} shows a typical energy spectrum obtained from
reconstructed hit energy before the $\phi$ symmetry calibration. It is shown
for a single channel in the HB (left) and the HF (right) calorimeters, along
with the values of the lower and upper thresholds used for these subdetectors.

The reconstructed energies are obtained from zero-suppressed events after
pedestal subtraction, and the corresponding mean noise level is zero.
The value of the lower threshold ($E_{\text{low}}$) depends on the subdetector
and is chosen to be well above (approximately 20 times) the RMS of the noise
distribution for a single channel, which is a few hundred \MeV.
The noise level is determined from pedestal data and is found to be a few
hundred \MeV. For the HB and HE calorimeters, the threshold is
4\GeV, whereas it is 10\GeV for the HF calorimeter.

The upper threshold ($E_{\text{high}}$)
ensures that statistical fluctuations in
the tails of the energy distributions do not influence the mean.
Its value also depends on the subdetector.
For $\Pp\Pp$ collision data, the threshold is 100 (150)\GeV for the HB
(HE and HF).

For each channel, the total energy between the thresholds
\etot is calculated from the observed energy
spectrum using:
\begin{linenomath}
\begin{equation}
\etot  =  \int_{E_{\text{low}}}^{E_{\text{high}}} \frac{\rd N(E)}{\rd E} E \rd E,
\end{equation}
\end{linenomath}
and the mean channel energy ($\langle \etot\rangle$) is defined as \etot
divided by the number of events used in the \etot calculation.
The scale factor is
calculated as the inverse of the ratio of $\langle \etot \rangle$ for that
channel to the mean $\langle \etot \rangle$
of all channels with the same $\ieta$ and depth, in each
iteration. These scale factors are then applied to the
energy measurement, and the whole process is repeated, including the selection
of the channels included in the determination of \etot through
the application of the energy thresholds. This procedure is repeated until
the mean change in the scale factor (over all channels) falls below a
convergence cutoff value.

\begin{figure}[htbp]
\begin{center}
  \includegraphics[width=.49\textwidth]{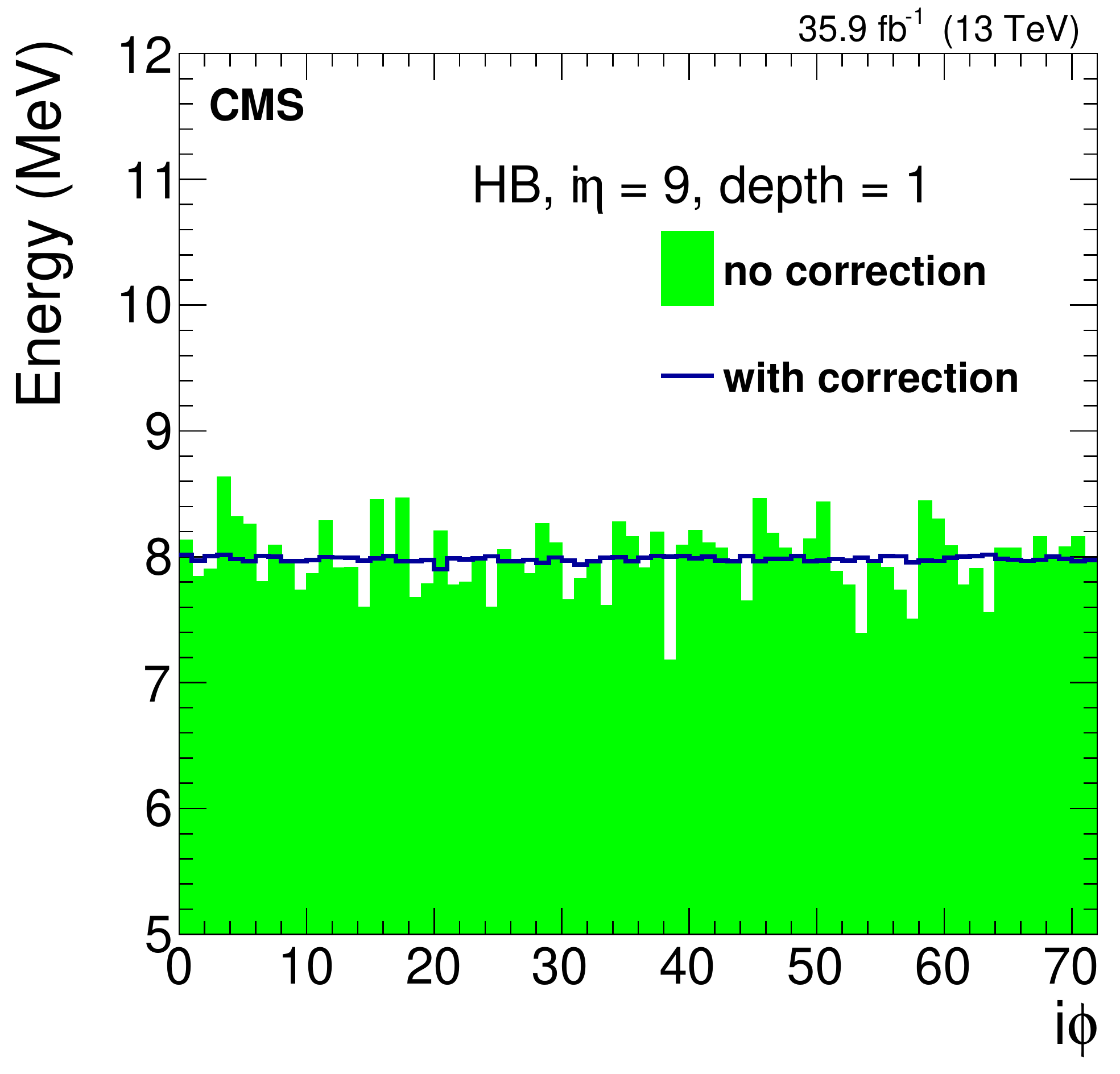}
  \includegraphics[width=.49\textwidth]{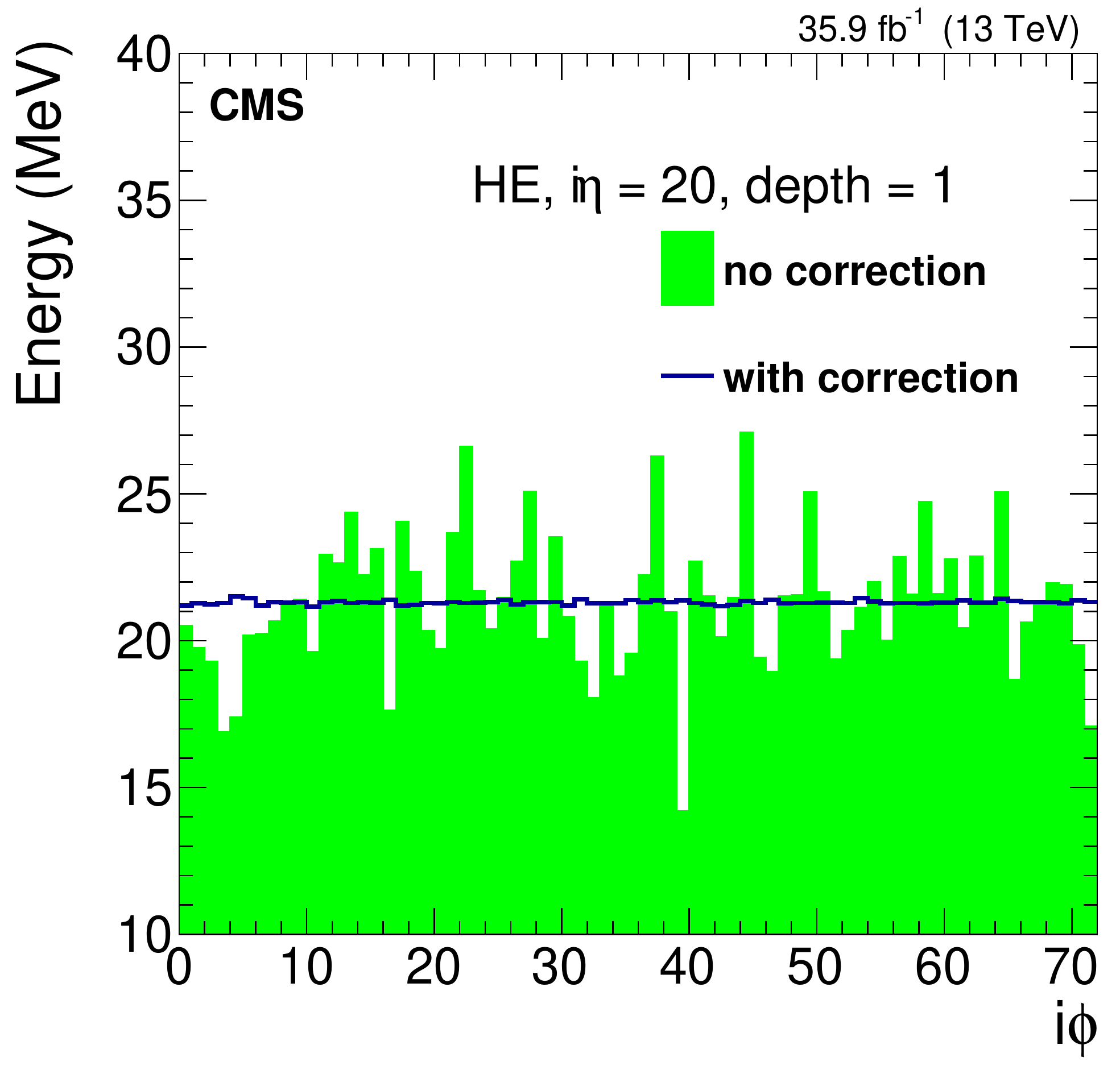}
 \end{center}
 \caption{Mean channel energy $\langle \etot \rangle$ for the iterative method
          measured before (solid histogram) and after (open histogram)
          correction as a function of $\iphi$ for a typical $\ieta$ ring of
          the HB ($\ieta = 9$, $\text{depth} = 1$, \cmsLeft) and of the HE
          ($\ieta = 20$, $\text{depth} = 1$, \cmsRight) calorimeters. Data
          triggered using electrons, photons, and muon are used in this
          calibration procedure.}
 \label{fig:iter1}
\end{figure}

Figure~\ref{fig:iter1} shows $\langle \etot \rangle$
as a function of $\iphi$ for two typical $\ieta$ rings for the HB and HE
calorimeters before and after the corrections.
The spread in the mean channel energy $\langle \etot \rangle$ is reduced
from 4.7 to 0.3\% for HB and from 6.2 to 0.2\% for HE.
The uncertainties in the scale factors are estimated from the statistical
uncertainties in $\langle \etot \rangle$ and the variation from
the last iteration, added in quadrature.

The statistical uncertainties, including the variations from the last iteration,
for the 2016 $\Pp\Pp$ collision data are of the order of 1\% for the HB,
between 0.1 and 1.0\% for the HE, and below 0.5\%
for the HF calorimeter channels,
depending on $\ieta$ and depth.

\subsection{Method of moments}\label{sec:moments}
The first two central moments of the energy distributions
are used to obtain the scale factors for each channel.
The main challenges in this method are (i)~the noise exceeds the size of a typical signal, which is a few MeV in the HB channels and a few tens of MeV in the HE channels, and (ii)~the variance of the noise distribution, 0.04--0.09\GeV$^2$, differs considerably from one channel to another.
Conditions are more favorable in the HF,
where the noise variance is less than the signal variance and
also less than the mean value of the signal.
Therefore, both the first and second moments for the HF calorimeter channels
are used in the determination of the intercalibration constants.

The analysis is done using MB events taken with a special trigger where
zero suppression is disabled in the HCAL readout.
The noise in each channel is measured separately by using an independent data
set taken when the LHC was not running and without any trigger requirements.
The measured noise distribution
is subtracted from the energy distribution after suitable
normalization (pedestal subtraction).

The scale factor obtained using the first moment is given by
\begin{linenomath}
\begin{equation}
{\mathrm{C}}_{\ieta,\iphi}  =
\frac{\langle E_{\ieta,\iphi}\rangle}{\frac{1}{N_{\phi}} \sum_{j\phi} \langle E_{\ieta,j\phi}\rangle},
\end{equation}
\end{linenomath}
where $N_{\phi}$ is the number of HCAL towers in a given $\ieta$ ring, and
\begin{linenomath}
\begin{equation}
\langle E_{\ieta,\iphi}\rangle  =  \langle E^{\text{signal}}_{\ieta,\iphi}\rangle + \langle E^{\text{noise}}_{\ieta,\iphi}\rangle
\label{eqn:one}
\end{equation}
\end{linenomath}
is the mean collected energy in the HCAL tower.
After pedestal subtraction, the data are consistent with
 $\langle E^{\text{noise}}_{\ieta,\iphi}\rangle = 0$ and
 Eq.~\eqref{eqn:one} becomes:
\begin{linenomath}
\begin{equation}
\langle E_{\ieta,\iphi}\rangle = \langle E^{\text{signal}}_{\ieta,\iphi}\rangle.
\end{equation}
\end{linenomath}

\begin{figure}[htbp]
\begin{center}
  \includegraphics[width=.49\textwidth]{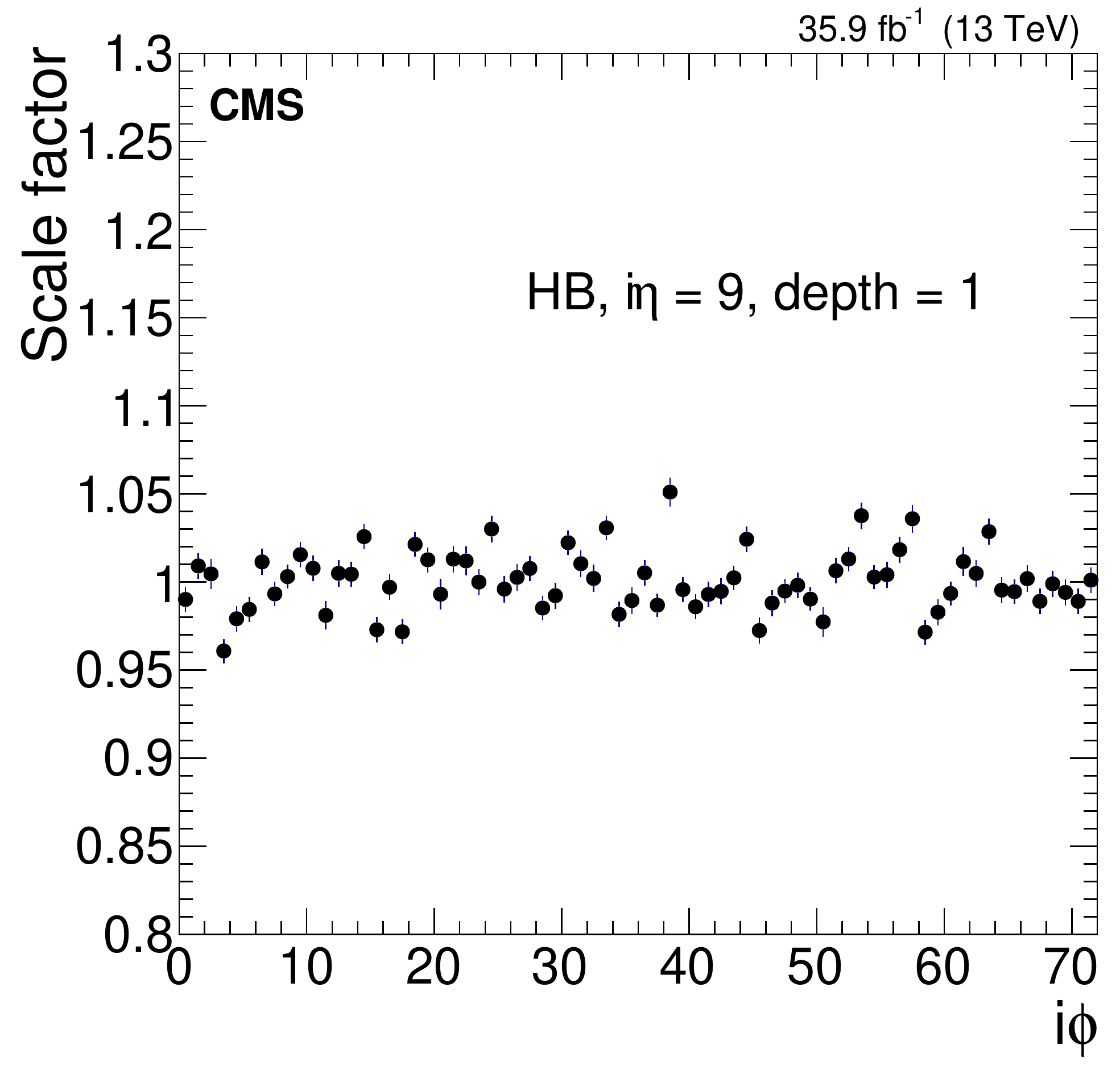}
  \includegraphics[width=.49\textwidth]{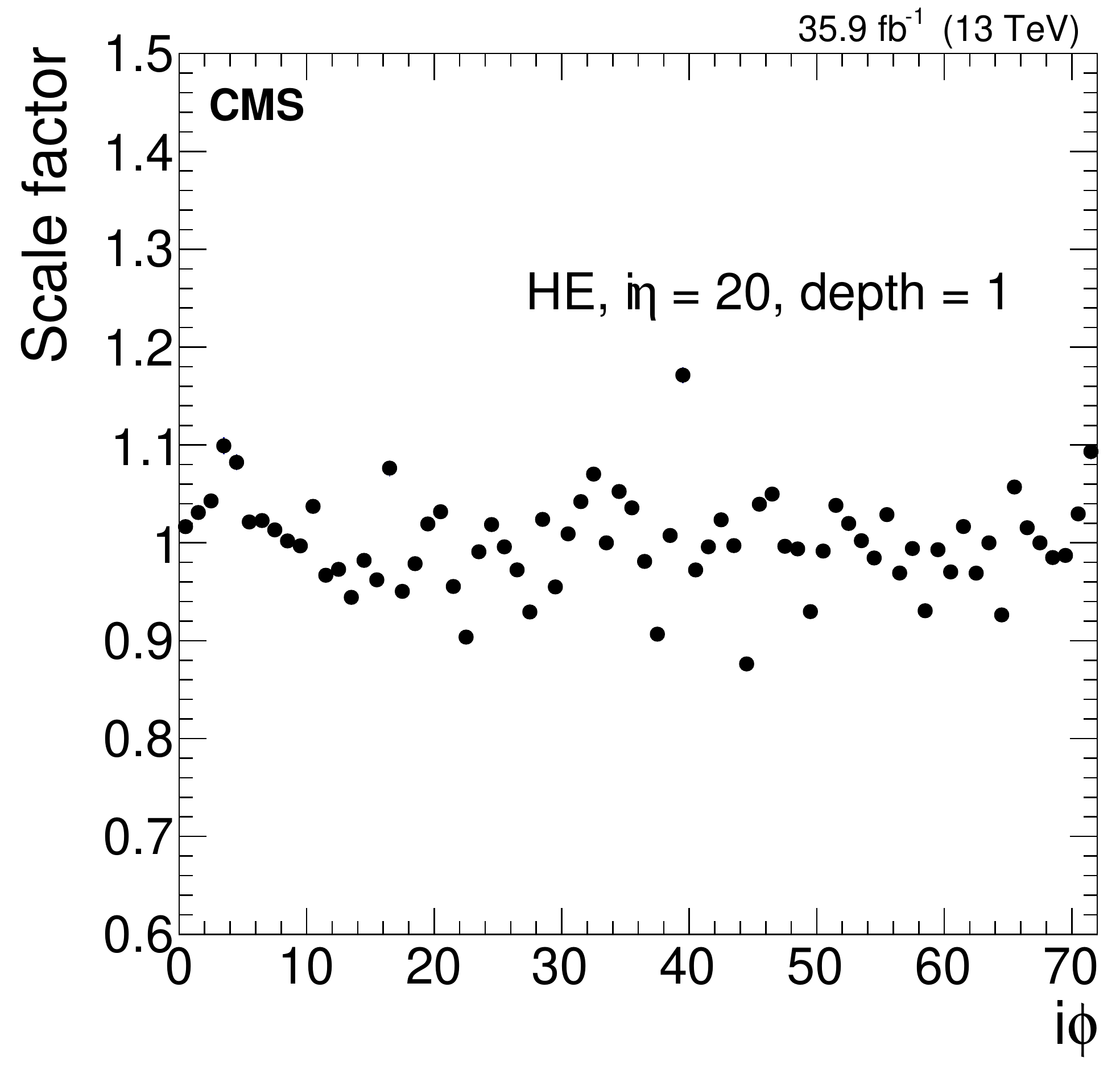}
 \end{center}
 \caption{Calibration scale factor obtained using the method of moments
          with the first moment of the energy distribution for one HB
          ($\ieta = 9$, $\text{depth} = 1$ \cmsLeft)and one HE
          ($\ieta = 20$. $\text{depth} = 1$ \cmsRight), as
          a function of $\iphi$. Only statistical uncertainties in the
          measurements are shown in these plots and the sizes of the vertical
          bars are typically smaller than the marker size.}
 \label{fig:momHF}
\end{figure}

The uncertainty in the estimation of the first moment is given by
\begin{linenomath}
\begin{equation}
\sqrt{\Delta^{2}\left(\langle E^{\text{signal}}_{\ieta,\iphi}\rangle\right) +
      \Delta^{2}\left(\langle E^{\text{noise}}_{\ieta,\iphi} \rangle\right)},
\end{equation}
\end{linenomath}
where $\Delta^{2}$ is the variance. It is dominated by the
uncertainty in the noise estimation. To achieve a precision
better than 2\% for channels in the middle of HB ($\abs{\ieta} = 1$),
a few tens of millions of events are required.
While the method is straightforward to perform, a large amount of data is required to use it effectively.
Figure~\ref{fig:momHF} shows the scale factors obtained for two specific towers
of HB and HE ($\ieta = 9$, $\text{depth} = 1$, and $\ieta = 20$,
$\text{depth} = 1$) using this method with data collected during 2016.

Estimation of the scale factor from the second central moment (variance)
is done after removing the noise by subtracting the variance of the noise from
the variance of the measured energy. The scale factor in this case is given
by:
\begin{linenomath}
\begin{equation}
\mathrm{C}_{\ieta,\iphi}  =
\sqrt{\frac{\frac{1}{N_{\phi}} \sum_{j\phi} \Delta^2R_{\ieta,j\phi}}{\Delta^2 R_{\ieta,\iphi}}},
\end{equation}
\end{linenomath}
where
\begin{linenomath}
\begin{equation}
\Delta^2R_{\ieta,\iphi}  =
\langle\Delta^2(E^{\text{signal}}_{\ieta,\iphi})+\Delta^2(E^{\text{noise}}_{\ieta,\iphi})\rangle -
\langle\Delta^2(E^{\text{noise}}_{\ieta,\iphi})\rangle.
\end{equation}
\end{linenomath}
Assuming no correlation between the noise and signal depositions in the
calorimeter,
\begin{linenomath}
\begin{equation}
\Delta^2R_{\ieta,\iphi}  =  \langle\Delta^2(E^{\text{signal}}_{\ieta,\iphi})\rangle.
\end{equation}
\end{linenomath}
The minimum sample size for achieving a 2\% uncertainty in the signal variance
is determined by
the residual noise contribution, and is of the order of a few million
events. The method based on the second moment requires substantially smaller
samples. Therefore, the second-moment method is used for the final results
for HB and HE, but the results are still sensitive to the noise level in
the channel, even when the noise levels are measured.
During 2016, the noise levels for channels in the
HF calorimeter were not measured, and the method based on the first moment is
used for those channels.

\begin{figure}[htbp]
\begin{center}
  \includegraphics[width=.49\textwidth]{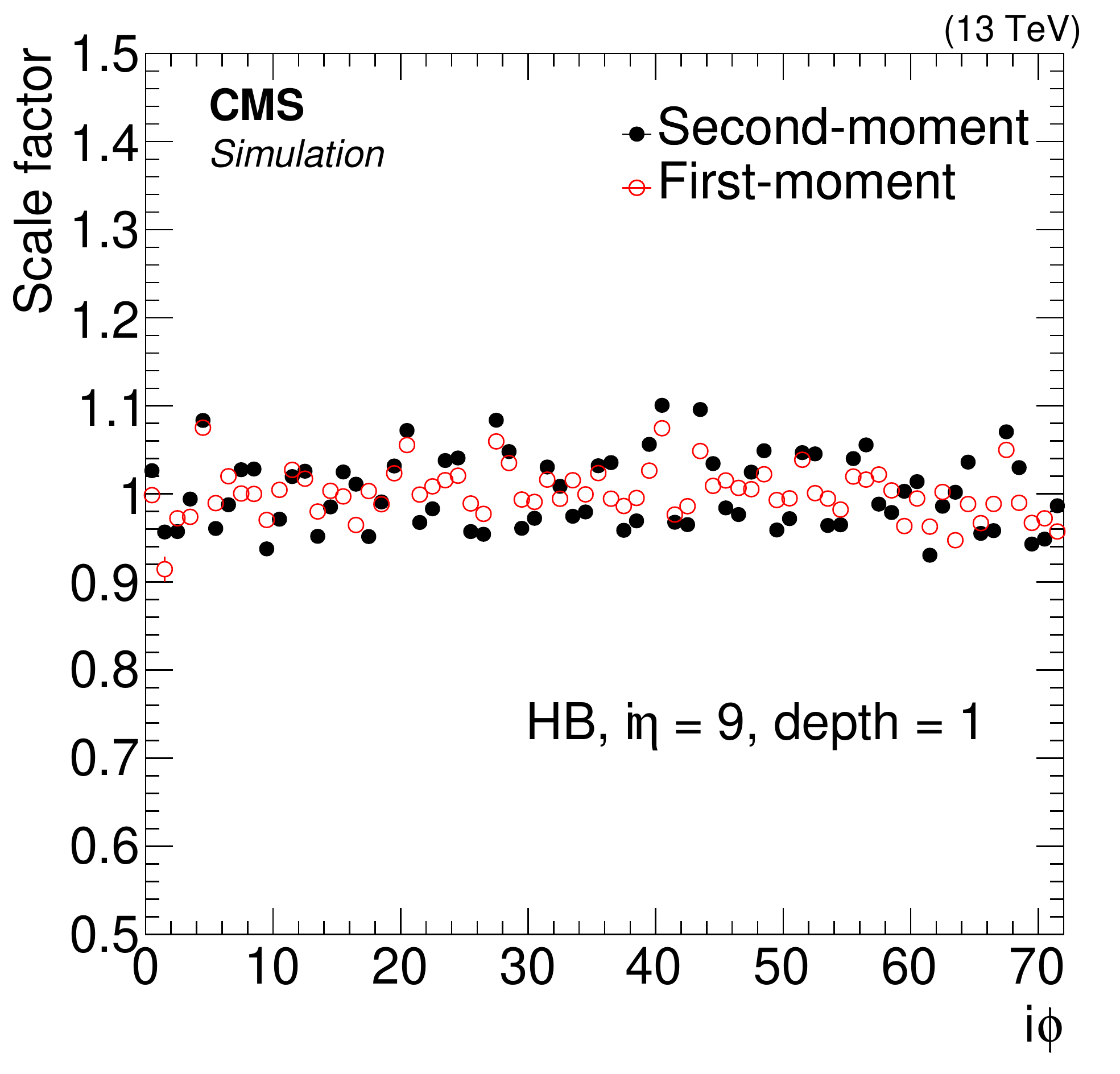}
  \includegraphics[width=.49\textwidth]{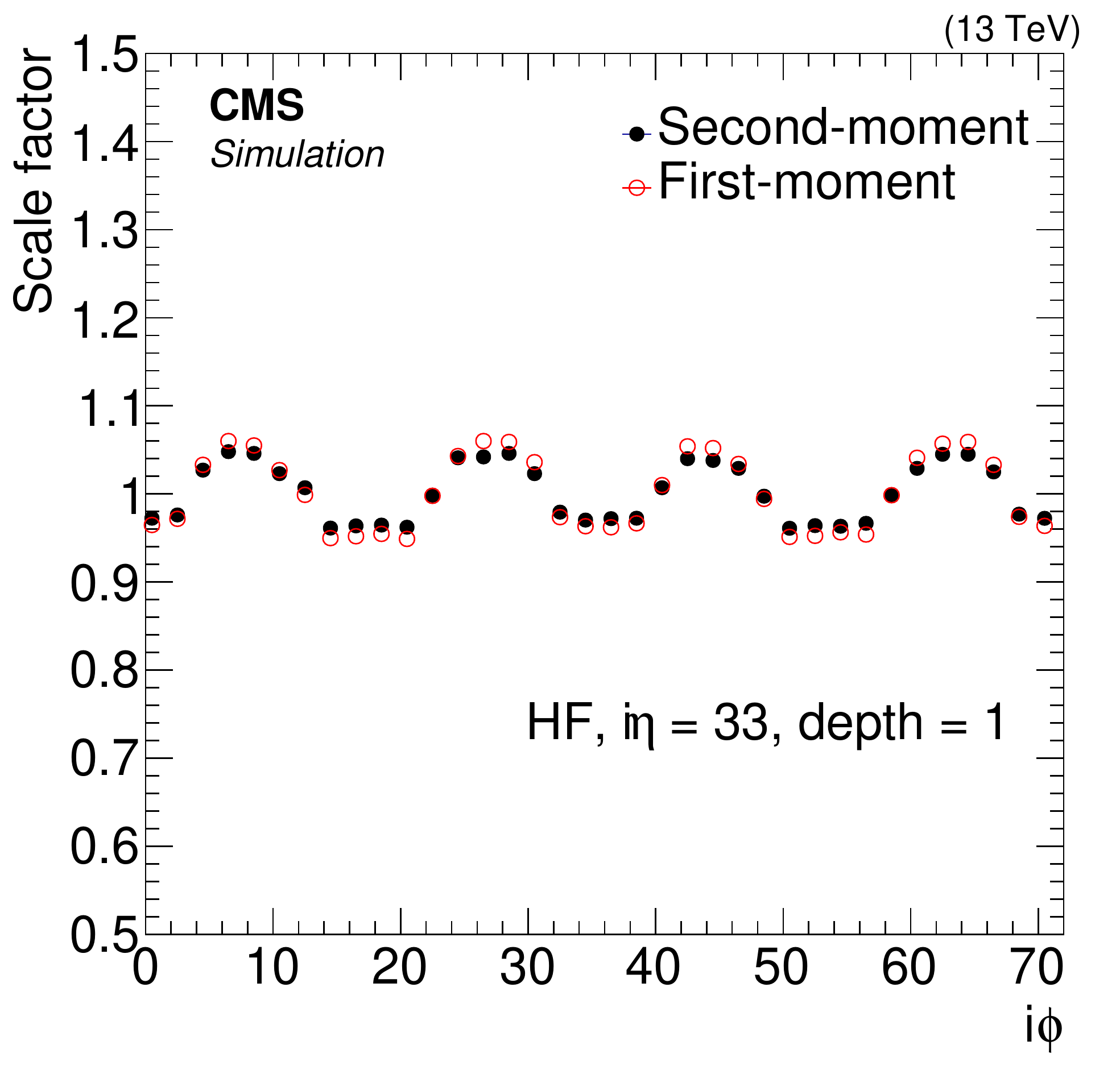}
 \end{center}
 \caption{Derived calibration scale factors that equalize the $\phi$ response
          in a simulated sample of minimum bias events for a channel in the HB
          ($\ieta = 9$, $\text{depth} = 1$, \cmsLeft) and HF ($\ieta = 33$,
          $\text{depth} = 1$, \cmsRight), as a function of $\iphi$ using two different
          methods: of first and of second moment. }
 \label{fig:momHBHF}
\end{figure}

Figure~\ref{fig:momHBHF} shows the effect of using the first- or
second-moment method on a simulated sample of MB events for the HB (left) and
HF (right).
The structure in the plot of HF scale factors as a function of $\iphi$ reflects
the geometry of the readout system (middle versus edge readouts) and of the
passive material between the calorimeter and the interaction point
(support structure and services for the detectors).
The two methods of moments (mean and variance) have different sensitivities to  the amount of material
in front of the calorimeter. For HB and HE, the
difference is typically $2.2\pm 0.1$\% on average, while for the HF
calorimeter this difference is somewhat smaller, $0.8\pm 0.1$\% on average.
This difference is assigned as a systematic uncertainty in the scale factor determination from the method of moments. An additional systematic uncertainty is assigned related to the determination of electronic pedestals using various sets of noncollision data. This uncertainty, which is negligible in the HB and varies in the 1.0--1.5\% range in the HE/HF, is added in quadrature to the uncertainty from the method of second moment.

\begin{figure}[htbp]
\begin{center}
  \includegraphics[width=.49\textwidth]{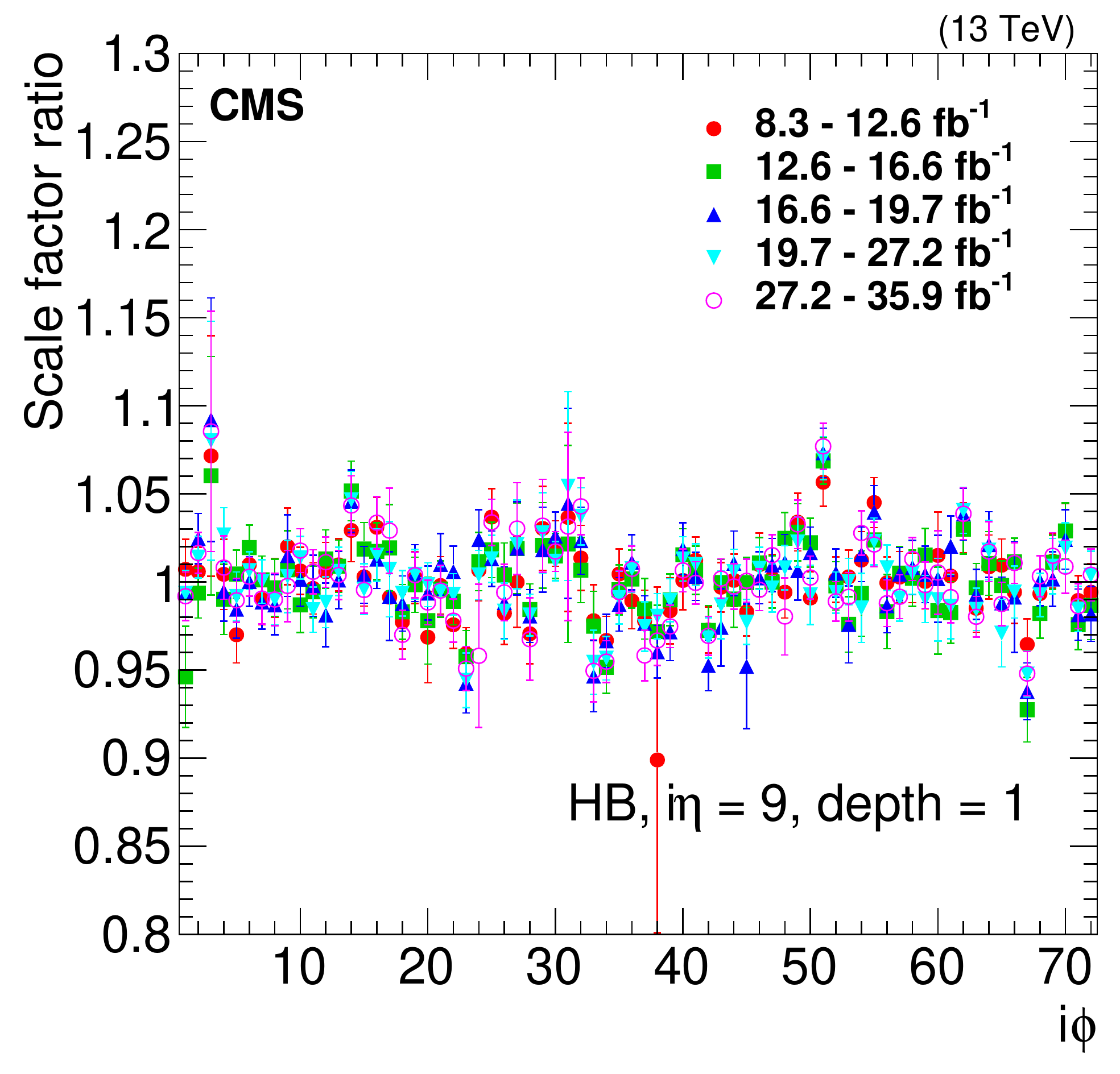}
  \includegraphics[width=.49\textwidth]{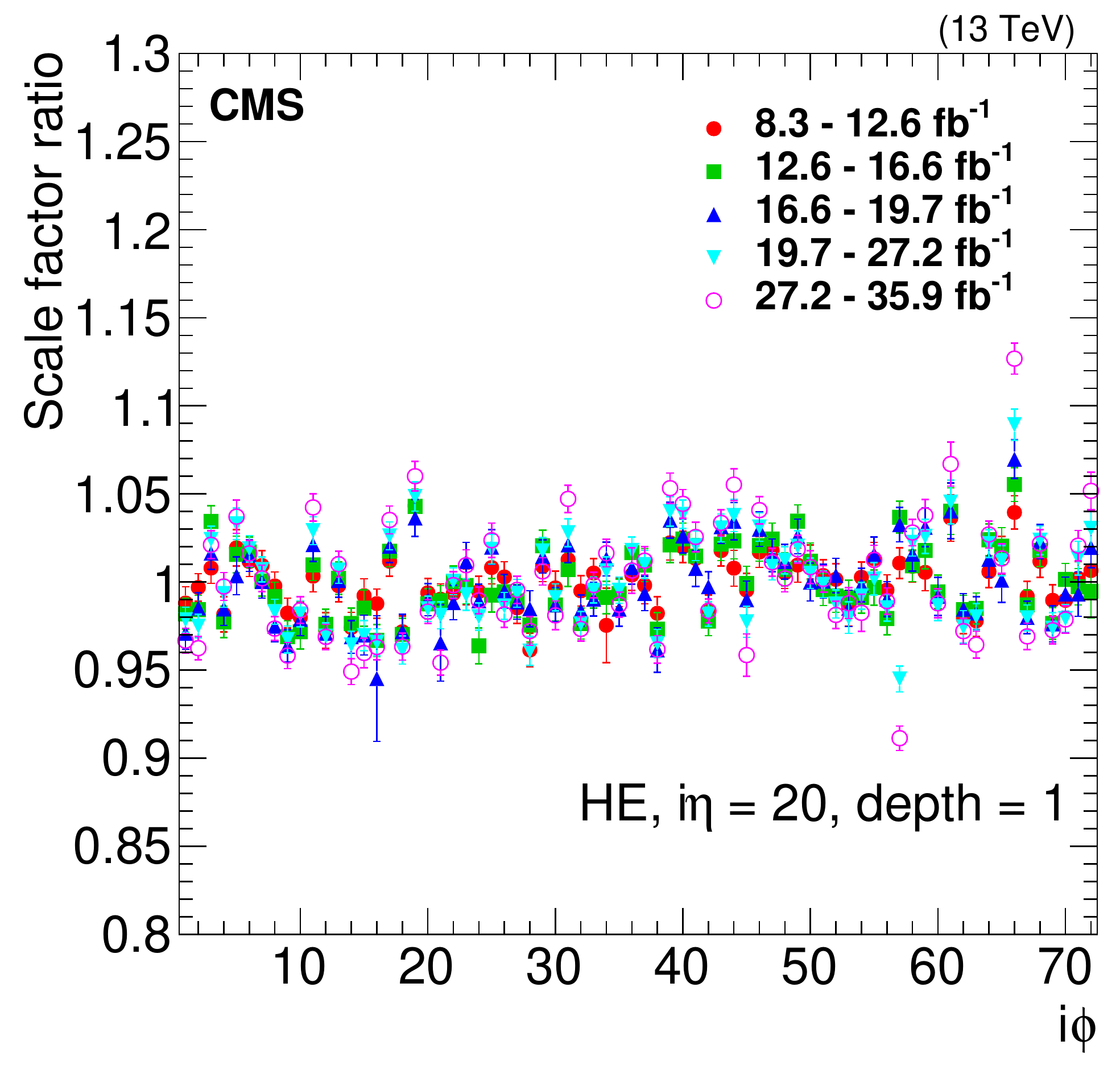}
 \end{center}
 \caption{Ratio of the calibration scale factors for a typical HB
          ($\ieta = 9$, $\text{depth} = 1$, \cmsLeft) and  HE ($\ieta = 20$,
          $\text{depth} = 1$, \cmsRight) channels as a function of $\iphi$, in
          different data taking periods, to that obtained in a sample
          corresponding to the first 8.3\fbinv\ of integrated luminosity
          for five additional data taking periods during 2016. Although the
          dataset is divided into five periods for the purpose of illustration,
          we end up with three sets of constants over the entire run of 2016.
          Only statistical uncertainties in the scale factors
          are shown.}
 \label{fig:momHBHE}
\end{figure}

Figure~\ref{fig:momHBHE} shows the ratios of scale factors obtained
in different portions of the 2016 data set
from the second-moment method for two representative channels: $\ieta = 9$,
$\text{depth} = 1$ in the HB, and $\ieta = 20$, $\text{depth} = 1$ in the HE.
The plots indicate the level of stability of the scale factors over
approximately six months of data taking.

\subsection{Combination of the two methods}\label{sec:combine}

\begin{figure}[htbp]
\begin{center}
  \includegraphics[width=.328\textwidth]{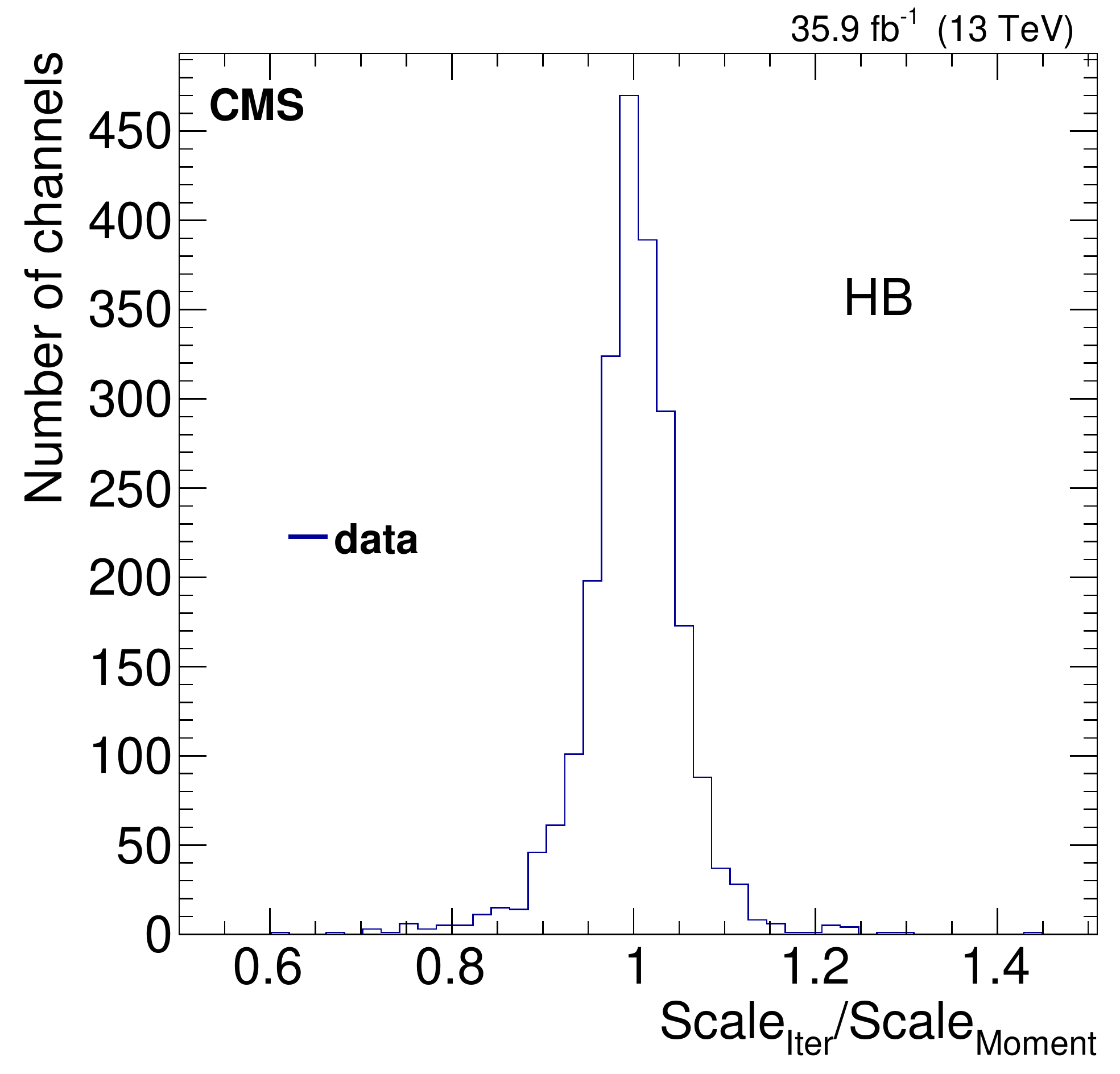}
  \includegraphics[width=.328\textwidth]{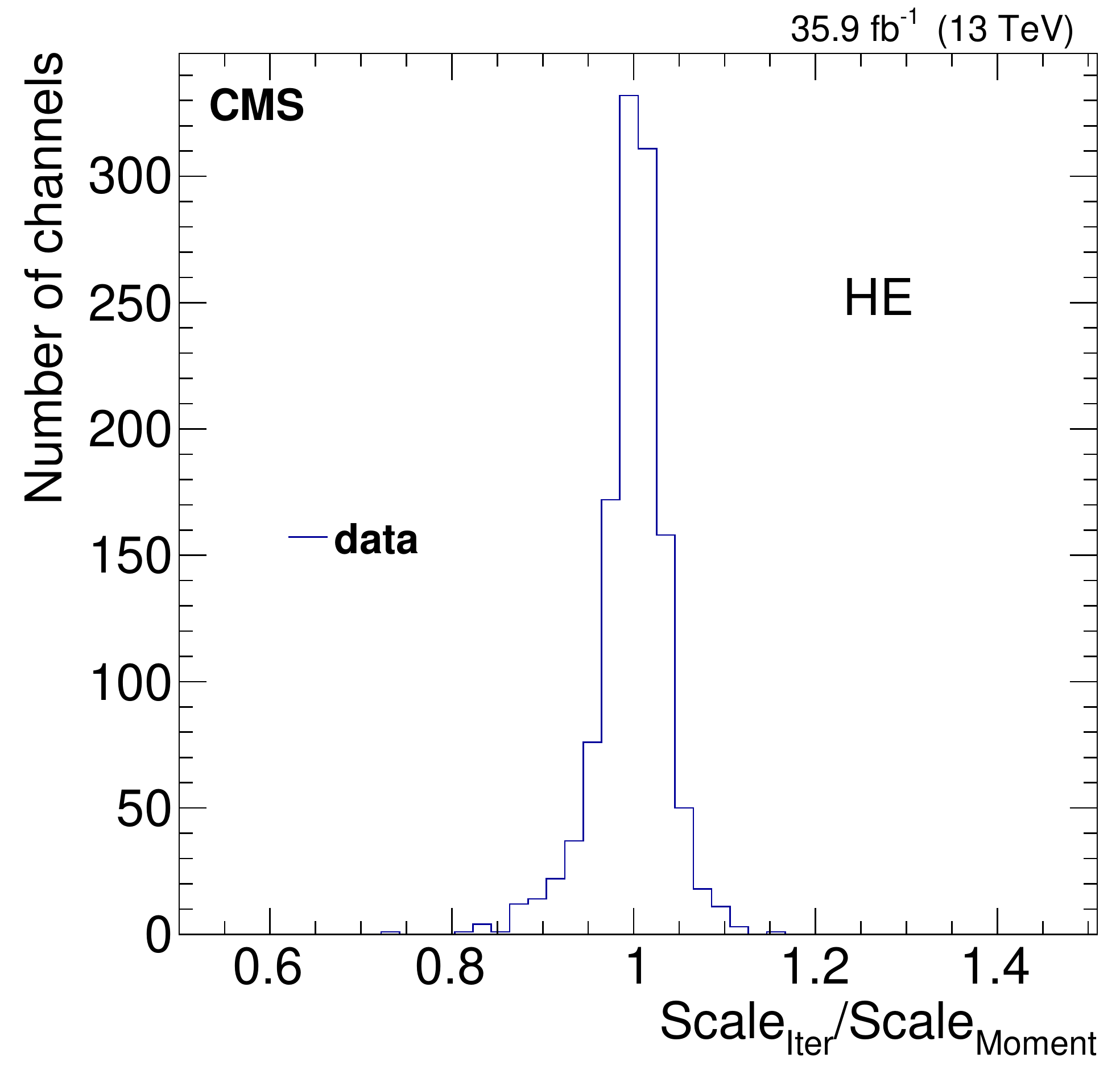}
  \includegraphics[width=.328\textwidth]{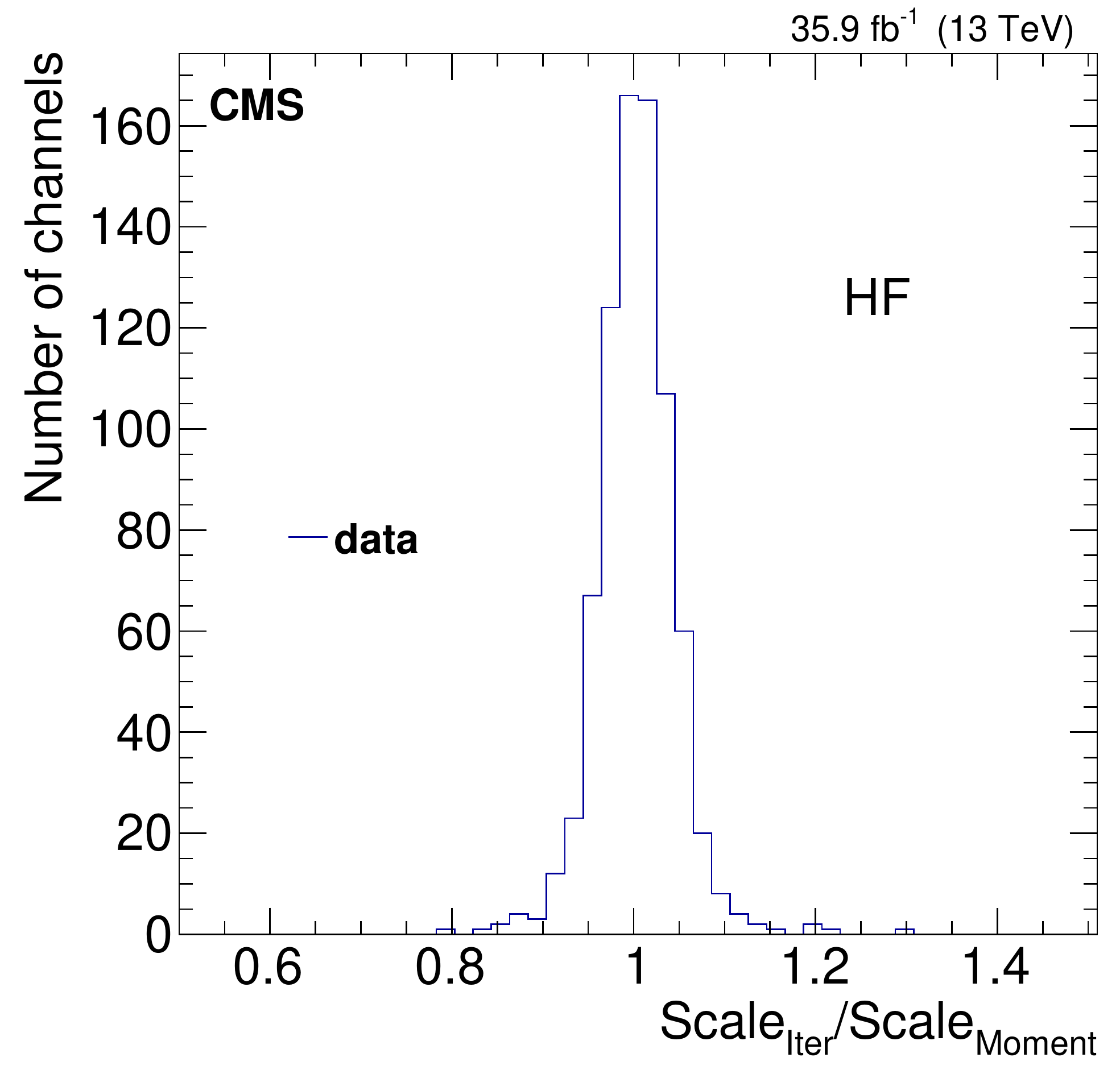}
 \end{center}
 \caption{The ratio of $\phi$ intercalibration scale factors obtained with the
          method of moments to that obtained with the iterative
          method for HB (left), HE (center), and HF (right). The mean, RMS
          and width from Gaussian fit to the distribution are $0.998\pm 0.001$,
          $0.058\pm 0.001$, $0.045\pm 0.001$ for the HB, $0.998\pm 0.001$, 
          $0.039\pm 0.001$, $0.030\pm 0.001$ for the HE, $1.004\pm 0.002$,
          $0.043\pm 0.001$, $0.037\pm 0.001$ for the HF.}
 \label{fig:phisym}
\end{figure}

Figure~\ref{fig:phisym} shows the ratio of the scale factors
measured using the two methods (iterative method over method of moments) for
the HB, HE, and HF calorimeters.
The two sets of measurements agree within 5\% as indicated from the
RMS of the ratio distributions.
The two sets of measurements agree, with the means of their ratios being
$0.998\pm 0.001$, $0.998\pm 0.001$, and
$1.004\pm 0.002$ for the HB, HE, and HF calorimeters, respectively.
The scale factor obtained from the method of moments has better precision
for low-energy depositions, whereas those from the iterative method are better
for high-energy depositions.
The statistical uncertainties from the two methods are comparable for the
HF calorimeter channels, whereas the iterative method gives smaller
uncertainties for the HB and HE calorimeter channels.
The uncertainty-weighted average of the scale factors from the two methods
is used as
the final scale factor for the HCAL $\phi$ intercalibration.
The arithmetic mean of the corrections is used when the statistical
uncertainties of both methods are below 1\%.  The weighted average
(with weight $w = 1/\sigma^2$, where $\sigma$ is the uncertainty in the
measurement) is used otherwise.
The systematic uncertainties in the interchannel calibration measured from the 
method of moments are estimated using (i)~the differences between constants 
found with the first and the second moments, which amount to 2--3\% for all 
three detectors: HB, HE, and HF; and (ii)~different sources of data: NZS 
(events recorded without zero suppression) and
single-muon trigger data, which yield values less than 1\%. The systematic 
uncertainties for the iterative method are estimated using (i)~different 
triggers: single-electron with $\pt>25\GeV$ and single-muon with $\pt
> 24\GeV$, which yield values below 1\%; and (ii)~different 
choices of the energy threshold in estimating \etot, which result in effects 
smaller than 3\%. The difference in the coefficients derived using the two 
methods provides a combination of systematic and statistical uncertainties, 
which is 4\% for the HB, and 3\% for the HE and HF.

\section{Absolute calibration using isolated tracks}\label{sec:isotk}

The energies of charged hadrons in the central region of the CMS detector
are measured by two independent detector systems: the trackers and the calorimeters.
The precise calibration of the tracker system can be transferred to the calorimeter
by comparing the two measurements.
Unlike the momentum measurement by the tracker, the hadronic energy response of the HCAL is not linear.
The nonlinearity is more pronounced at lower energies~\cite{TB06}.
The goal of the HCAL absolute energy calibration is to set the
relative energy scale to unity for 50\GeV\ charged hadrons that
do not interact hadronically in the ECAL.
In practice, the calibration is done with
tracks of momentum between 40 and 60\GeV.
The dominant systematic uncertainties are due to the contamination of the
calorimeter energy from other hadrons, produced either in the same interaction
that produced the isolated charged-hadron candidate or in pileup interactions.

The data used in this method come from two sources.
The first sample is selected using a trigger designed for this analysis.
At L1, the event is required to contain at least one jet
with  $\pt > 60$\GeV. At the HLT, an isolated track with
an associated energy in the ECAL below 2\GeV\ is required.
The isolation calculation utilizes information from tracks after extrapolation to the calorimeter surface.
The second sample uses events from the full, varied suite of CMS triggers.
Events are selected offline using a filter that requires
a track that satisfies very loose isolation criteria, has
associated energy in the ECAL less
than 2\GeV, and momentum higher than 20\GeV.

Events are required to have at least one well-reconstructed
primary vertex~\cite{TRK-11-001} that is
close to the nominal interaction point, with
$r(\equiv\sqrt{x^2+y^2}) < 2$\cm and $\abs{z} < 15$\cm.
Tracks considered as candidate isolated hadrons
are required to be associated with the primary vertex and to
satisfy quality requirements.
Their impact parameters are required to be close to the primary
vertex in the transverse ($xy$) plane ($\Delta r < 200$\mum)
as well as along the beam axis ($\Delta z < 200$\mum).
The $\chi^2$ of the track fit per degree of freedom
is required to be less than 5,
and the number of tracker layers used in the momentum measurement to be greater
than 8.
To confidently  select tracks that have not interacted
before reaching the calorimeter surface,
tracks with missing hits in the inner and outer layers of the tracker are
rejected.

The analysis uses isolation in a cone around the track
to reduce contamination from neighboring
hadrons, and to have a more accurate estimation of the hadron energy.
The cone algorithm clusters energy based on the linear distance from the
extrapolated track trajectory through the HCAL.
For each HCAL tower, the distance between two points is determined.
The first point is the intersection of the extrapolated track trajectory with
the front face of the HCAL. The second point is the intersection of the
tower axis (the straight line joining the center of the CMS and the center of
the tower) with the plane perpendicular to the extrapolated track trajectory.
If this distance is smaller than the radius of a circle on the surface of HCAL
($R_{\text{cone}}$), the energy from the HCAL tower is included in the cluster.
The signal is measured using $R_{\text{cone}}$ of 35\cm, which contains on
average more than 99\% of the energy deposited by a 50\GeV hadron.

The ECAL has a depth of approximately one interaction
length and therefore more than half of the hadrons undergo inelastic
interactions before reaching the HCAL.
These hadrons are not used for
calibration and are rejected by requiring the energy deposited within a cone
of radius 14\cm around the impact point for the ECAL to be less than 1\GeV.
This requirement also removes a large fraction of
hadron candidates near  a neutral particle, which deposit energy
in the signal cone that would otherwise contaminate the measurement.

To further reduce contamination due to neighboring particles,
an isolation requirement based on tracking information is used.
The trajectories of
all charged particles are propagated to the ECAL surface.
Hadron candidates are vetoed if there are
additional tracks above a momentum threshold  impacting the calorimeter
surface within a
circle  of radius 64\cm around the candidate's impact point.
The momentum
threshold depends on the desired selection efficiency and purity.
Two threshold values are used:
a 10\GeV\ loose isolation requirement is used for the main analysis, whereas a
2\GeV\ tight isolation requirement is used in the assessment of systematic
uncertainties.
Although a stringent requirement on the isolation from neighboring charged
particles would reduce the uncertainty caused by contamination of the signal energy
from other nearby hadrons,
the large number of pileup interactions at high instantaneous luminosity
can result in a large amount of unrelated energy in the signal
cone in the endcap region of the HCAL.
The loose isolation requirement increases
the efficiency for the track selection in the presence of {\pu}.

The calorimeter response is defined as the ratio
\begin{linenomath}
\begin{equation}
E_{\mathrm{HCAL}}/(p_{\mathrm{track}} - E_{\mathrm{ECAL}}),
\end{equation}
\end{linenomath}
where  $E_{\mathrm{HCAL}}$ is the signal region energy of the HCAL cluster,
$E_{\mathrm{ECAL}}$ is the energy deposited in the ECAL in a cone of radius
14\cm around the impact point of the track,
and $p_{\mathrm{track}}$ is the momentum of the track.
To mitigate the contamination from {\pu}, we apply a {\pu} correction,
discussed below.
The most probable value (MPV or mode) of the response is extracted using an iterative
two-step fitting procedure to a Gaussian function. The two fits are
performed  in the intervals $\pm 1.5 \sigma_{\mathrm{Tot}}$ and
$\pm 1.5 \sigma_{\mathrm{Fit}}$ around the mean of the distribution or
the fitted mean, where $\sigma_{\mathrm{Tot}}$ is the sample RMS
and $\sigma_{\mathrm{Fit}}$ is the width determined from the first fit.

The contribution from {\pu} is subtracted on an event-by-event basis by
measuring the energy within an annular region beyond the signal cone.
When there is a large amount of {\pu} energy near the cone,
a relatively high {\pu} contribution is expected inside the cone.
If so, the true
particle energy is lower than that reconstructed in the cone.
If the energy deposit outside the cone is caused by the particle itself, the
true particle energy will be higher than the reconstructed energy in the cone.
The track under consideration and the {\pu} originate from independent collisions
within the same bunch crossing.
From a study using simulated single isolated high-\pt\ pion
events, the {\pu} contribution is related to
the energy in a region around $R_{\text{cone}}$ with an annular
radius of $+$10\cm.
It does not depend on the track momentum, and rather depends
on the ratio of the energy in the cone to the track momentum.
The corrected energy $E_{\text{cor}}$ is
calculated on an event-by-event basis using Eq.~\eqref{eq:corforpu},
\begin{linenomath}
\begin{equation}
\begin{aligned}
\label{eq:corforpu}
 E_{\text{cor}} & =  E\left( 1 + a_{1}\frac{E}{p}\left(\frac{\Delta}{p}
              + a_{2} \left( \frac{\Delta}{p} \right)^2\right)\right), \\
              (a_1, a_2 ) & =  \begin{cases}
              (-0.35, -0.65) & \text{for}~\abs{\ieta} < 25, \\
              (-0.35, -0.30) & \text{for}~\abs{\ieta} = 25, \\
              (-0.45, -0.10) & \text{for}~\abs{\ieta} > 25, \\
              \end{cases}
\end{aligned}
\end{equation}
\end{linenomath}
where $E$ is the energy in the signal region cone  $R_{\text{cone}} = 35$\cm
around the impact point of the selected isolated track, $p$ is the track
momentum, and $\Delta$ is the energy deposit in the annular region around
the signal cone.
The values for the constant depend on $\abs{\eta}$ because of the tracker
coverage and the $\abs{\eta}$ dependence of the
{\pu} particle energies.
The values of the coefficients $a_1$ and $a_2$
are extracted using the dependence of the response on the ratio
$\Delta/p$ by minimizing the difference between the mean corrected
response for simulated samples that are processed two ways: with and
without {\pu}.
The {\pu} scale factors are derived using single pion
simulated samples with and without {\pu}, and cross-checked with
independent single pion simulated samples with {\pu}.

\begin{figure}[htbp]
 \begin{center}
  \includegraphics[width=.49\textwidth]{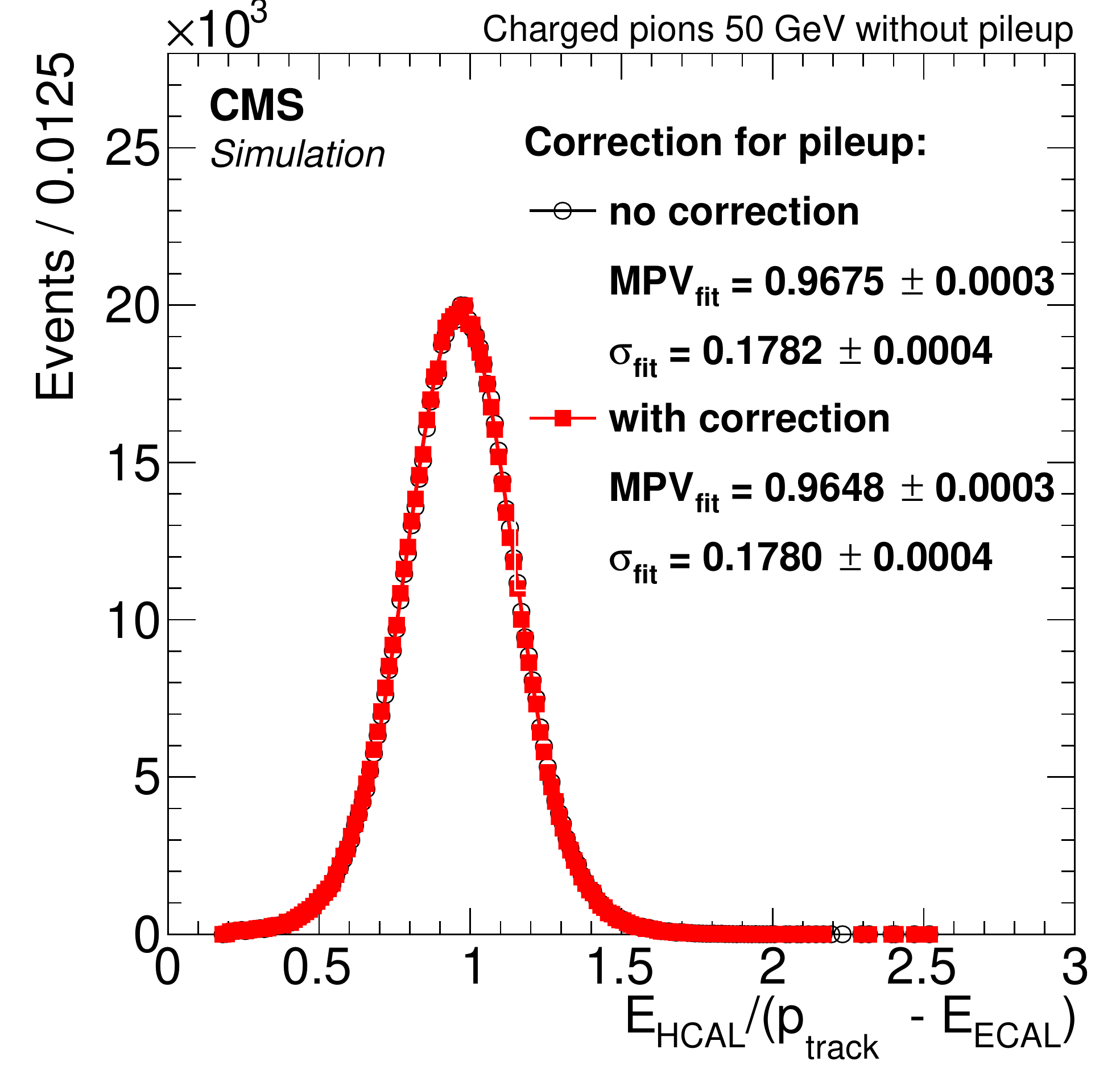}
  \includegraphics[width=.49\textwidth]{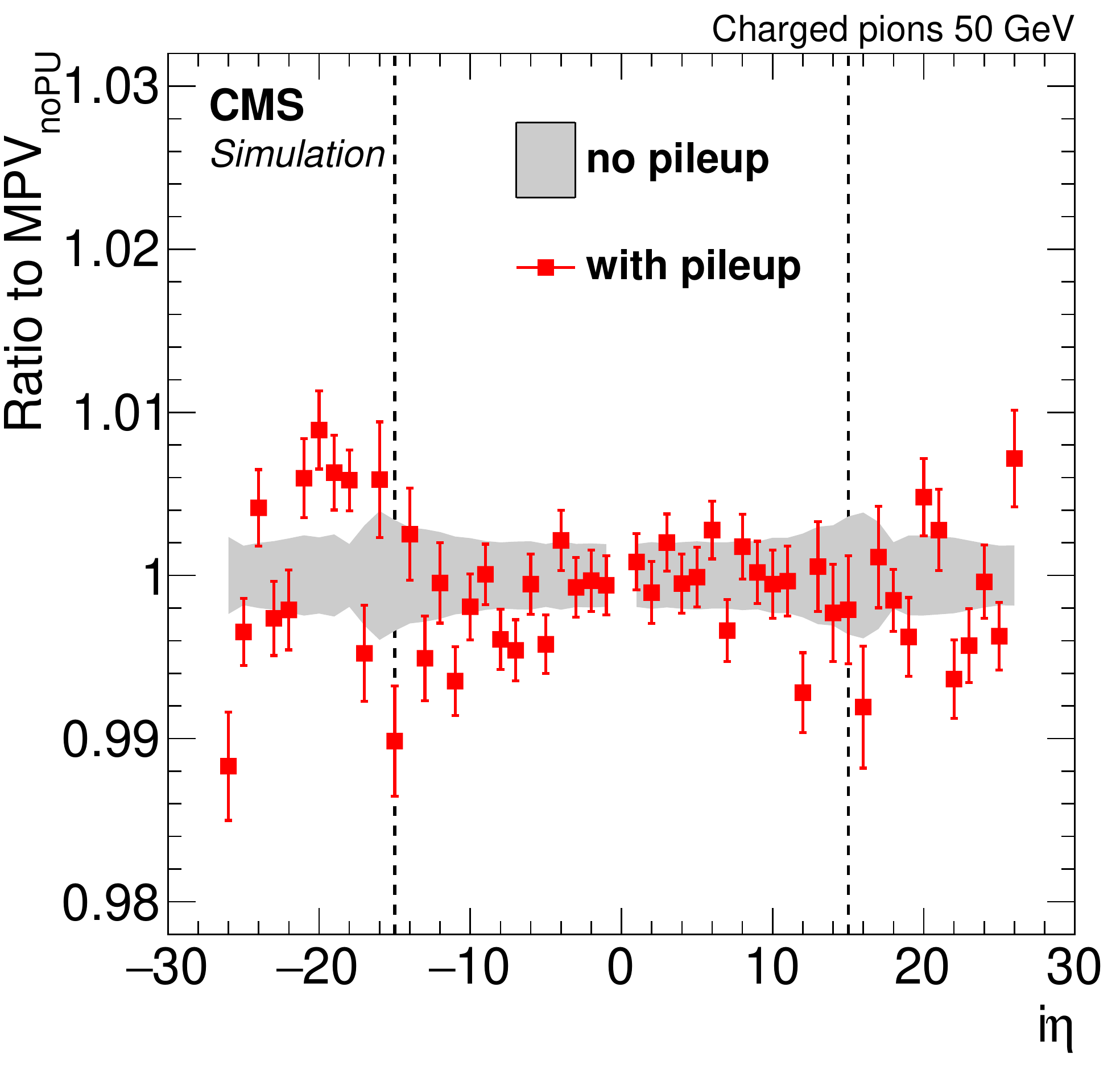}
 \end{center}
 \caption{(\cmsLeft) Distribution of the energy response in a simulated sample
          of single isolated high-\pt\ pions
          without {\pu} when the corrections for {\pu} have (red squares) or
          have not (black circles)  been applied.
          The plot also shows the results of Gaussian fits.
          (\cmsRight) The ratio of the mode of the response distribution for a
          simulated pion sample with {\pu}, with loose charged-particle
          isolation, and the correction for {\pu} applied (red squares) to
          the mode from the sample without {\pu}. The uncertainties in the mode
          without {\pu} are shown with the gray band. Only statistical
          uncertainties are included. The dashed vertical lines
          show the boundaries between the barrel and the endcaps.}
\label{fig:pu_cor}
\end{figure}

Figure \ref{fig:pu_cor} (left) shows the response distribution for the sample
of simulated pions without pileup before and after application of residual
energy scale corrections using the isolated track
calibration technique. The bias in the mode of the energy distribution for the
simulated pion sample that is caused by a {\pu} correction is less than 0.3\%.
The right plot in Fig.~\ref{fig:pu_cor} shows the ratio of the modal value
from a simulated single pion sample with {\pu} to that from a sample without
{\pu}.
The pions in the sample with {\pu} are required to satisfy the isolation criteria, and
their analysis utilizes the {\pu} correction technique described above.
The modes agree to within 1\%
for the entire calorimeter, and within 0.5\% in the barrel region.

The calibration method utilizes an iterative approach. At the $m$-th iteration,
the new scale factor $c_i^{(m+1)}$ is calculated using:
\begin{linenomath}
\begin{equation}
\label{eq:iterMinus}
c_i^{(m+1)} = c_i^{(m)} \left( 1 - \frac{\sum_j w_{ij}^{(m)}\left(\frac{E_{j}^{(m)}}{p_{j} - E_{j,\text{ECAL}}} - \text{RR}\right)}{\sum_j  w_{ij}^{(m)}} \right),
\end{equation}
\end{linenomath}

\noindent
where the sum is over events that contribute to the towers at the $i$-th
$\ieta$ ring, RR is the reference to which the mean response is
equalized, $E_{j,\text{ECAL}}$ is the measured energy in the ECAL cluster around
the track $j$, $p_{j}$ is the track momentum, $w_{ij}^{(m)}$ is the weight of the
particular tower with measured energy $e_{ij}$
in the cluster energy $E_{j}^{(m)}$:
\begin{linenomath}
\begin{equation}
 w_{ij}^{(m)}  =  \frac{c_{i}^{(m)} e_{ij}}{E_{j}^{(m)}}, \quad \quad
E_{j}^{(m)}  =  \sum_{i=1}^{n_j} c_{i}^{(m)} e_{ij}.
\end{equation}
\end{linenomath}

It follows from Eq.~\eqref{eq:iterMinus} that the iterative
procedure results in equalization of the mean response of the detector around
the value RR, which equals 1 by default. If the most probable value for the
sample, mode$_{\text{sample}}$,  differs from the sample mean,
mean$_{\text{sample}}$, the reference response is set to
$\text{RR} = \text{mean}_{\text{sample}}/\text{mode}_{\text{sample}}$.
The formulation in Eq.~\eqref{eq:iterMinus} makes the procedure stable with
respect to fluctuations of $E_{j}^{(m)}$.

The statistical uncertainty in the scale factor $\Delta c_{i}^{(m+1)}$ is
estimated from the measured RMS of the response distribution
$\Delta R_{i}^{(m)}$ for the subsample used for the $i$-th subdetector:
\begin{linenomath}
\begin{equation}
\Delta c_{i}^{(m+1)} = \Delta R_{i}^{(m)} \frac{\sqrt{\sum_{j}{(w^{(m)}_{ij})^{2}}}}{\sum_{j}{w^{(m)}_{ij}}}.
\end{equation}
\end{linenomath}

The procedure is iterated until the difference between the scale factors
in subsequent steps becomes three times smaller than the statistical
uncertainty.

\begin{figure}[htbp]
 \begin{center}
  \includegraphics[width=.328\textwidth]{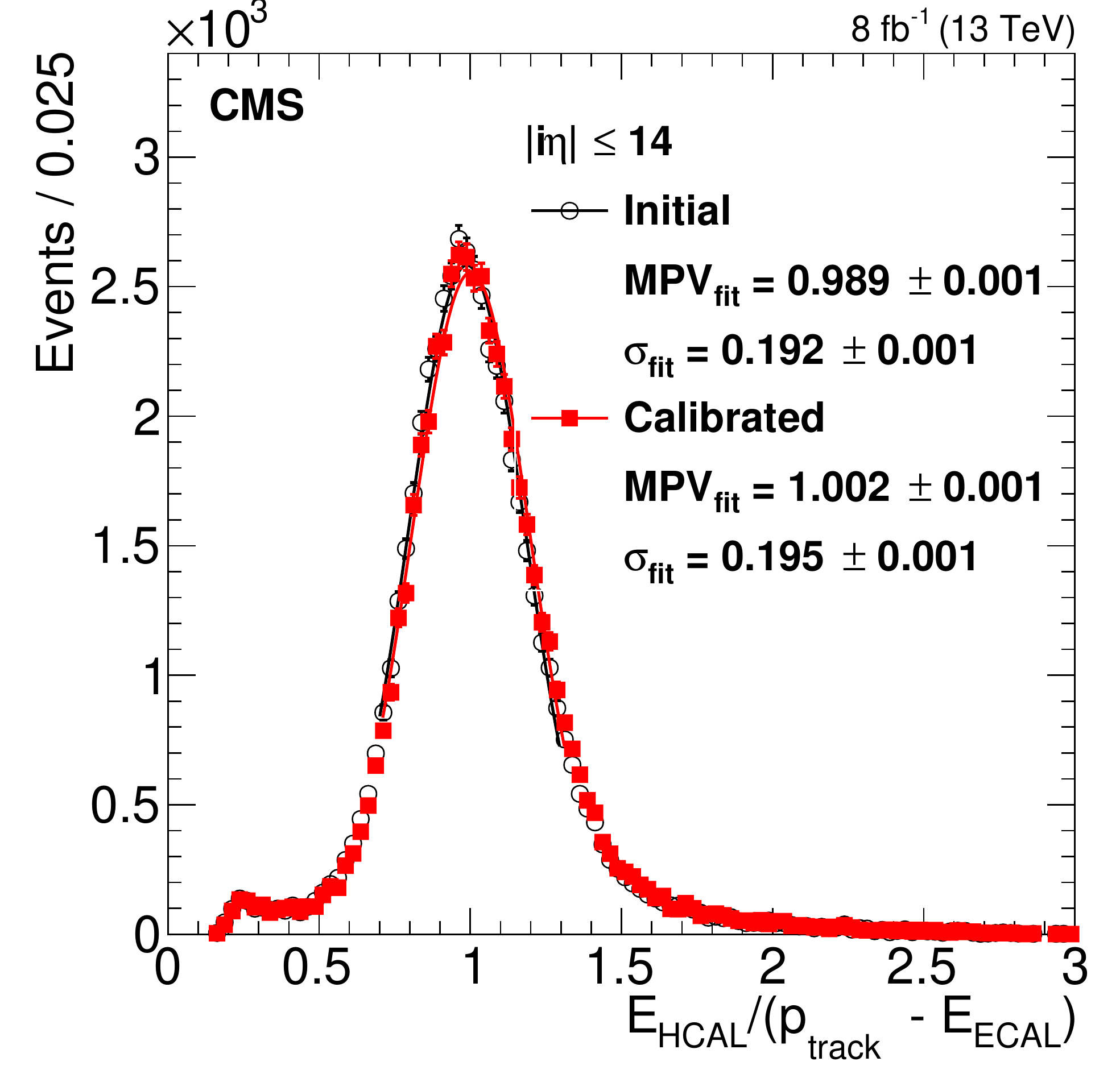}
  \includegraphics[width=.328\textwidth]{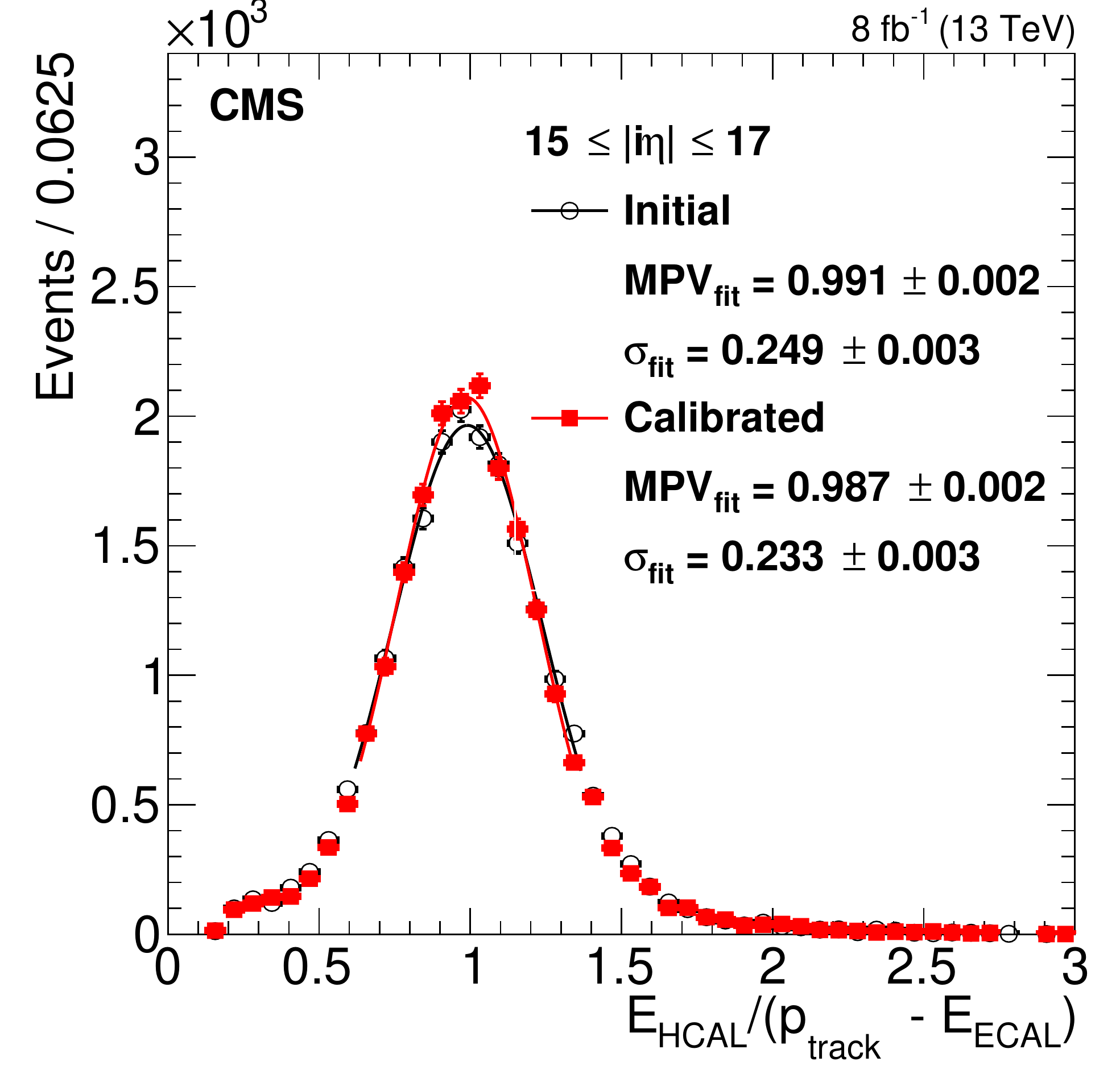}
  \includegraphics[width=.328\textwidth]{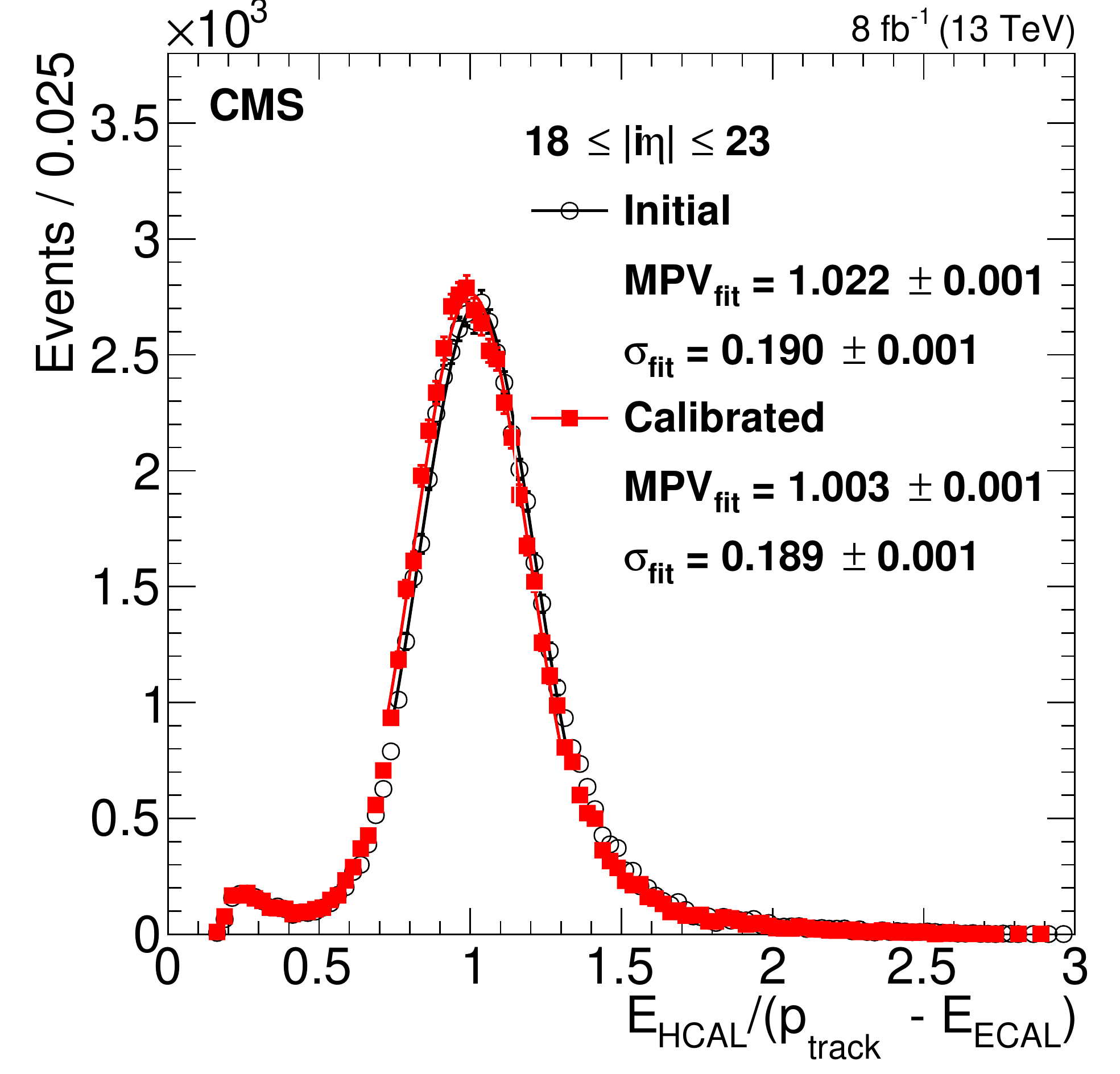}
 \end{center}
 \caption{Response distributions for pions from 2016 data in three different
          $\eta$ regions, $\abs{\eta} \leq 1.22$ (left),
          $1.22 \leq \abs{\eta} \leq 1.48$ (middle), and
          $1.48 \leq \abs{\eta} \leq 2.04$ (right), with loose
          charged-particle isolation criterion: initial (black circles) and
          after convergence (red squares). Only statistical uncertainties
          are shown on the data points.}
 \label{fig:e2p_2016DEG}
\end{figure}

\begin{figure}[htbp]
 \begin{center}
  \includegraphics[width=.6\textwidth]{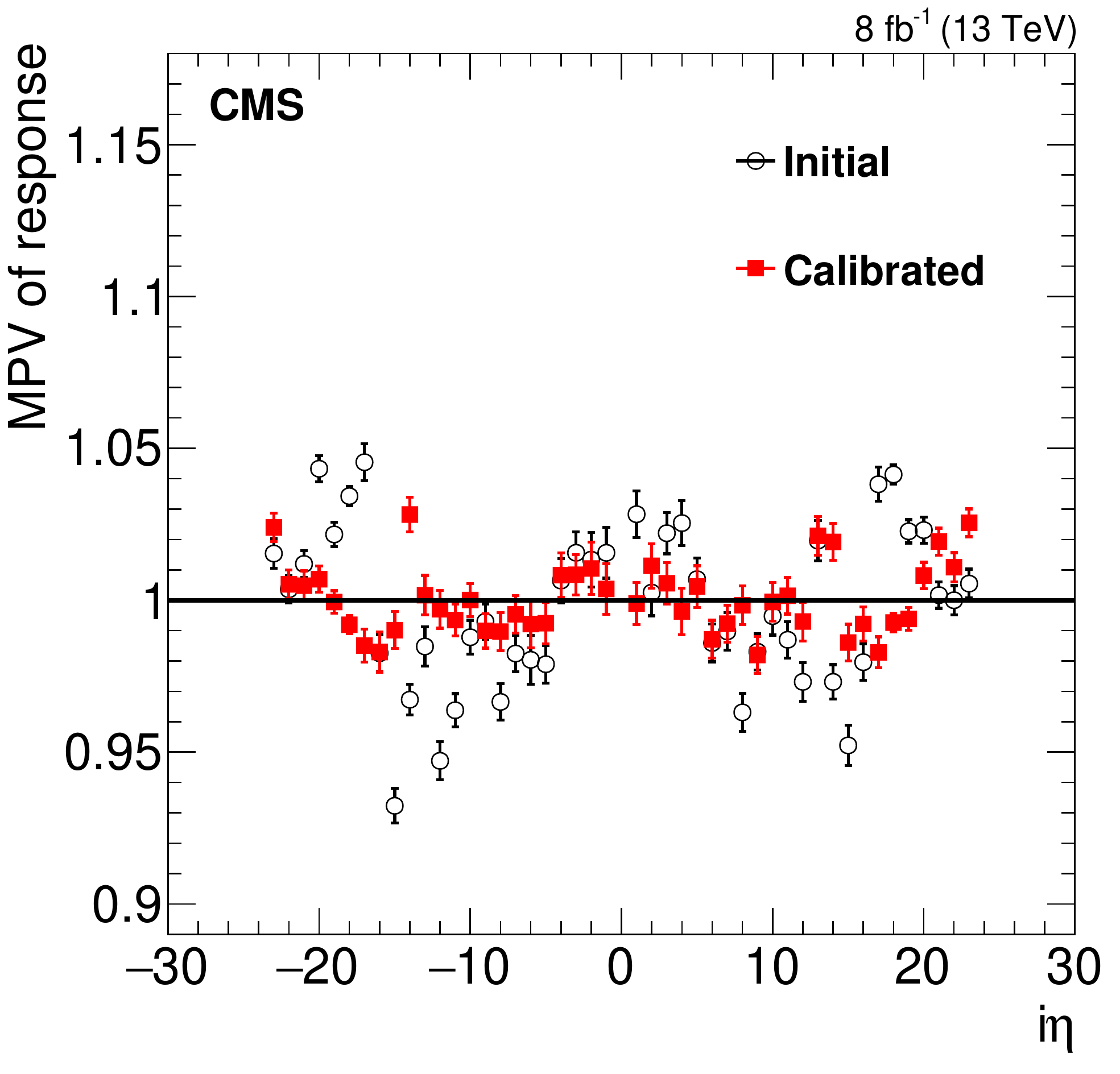}
 \end{center}
 \caption{Modes of the response with their statistical uncertainties versus
          $\ieta$ from the 2016 data sample before (black circles)
          and after convergence (red squares). The loose charged-particle
          isolation constraint is applied.}
 \label{fig:mpv_2016DEG}
\end{figure}

The calibration procedure is applied to the 2016 collision data to obtain
scale factors for each $\ieta$ ring through $\ieta = 23$. Scale factors
for rings beyond $\ieta = 23$ in the HE are obtained by extrapolating these results.
We note that the determination of the scale factors cannot be extended all the way to the boundaries of the tracker coverage, because the tracks used in the calibration are required to satisfy isolation criteria with respect to other particles, so the entire isolation cone is required to be within the tracker acceptance.
Different criteria for selection of isolated hadrons and different methods for
the pileup correction are tested, and the resulting correction
factors for each data taking period are compared.
The initial and resulting (after convergence) response distributions are
shown in Fig.~\ref{fig:e2p_2016DEG} for three HCAL $\eta$ ranges.
The resulting equalization of the mode is shown in
Fig.~\ref{fig:mpv_2016DEG}. Equalization within $\pm 2.5$\% is achieved
with the iterative procedure for subdetectors up to $\abs{\ieta} = 23$.

The statistical uncertainty in the scale factor, as obtained from data,
is typically below 2\%. 
The uncertainty in the scale factor is computed from the differences obtained 
(i)~using tight and loose isolation criteria (estimated to be $<$1\% for 
$\abs{\ieta}<15$, and $<$2\% for $\abs{\ieta}> 14$); 
(ii)~using simulated isolated high-\pt pion samples with and without pileup 
(estimated to be less than 0.1\% over the entire \ieta range); and 
(iii)~using the true momenta versus the measured momenta of the charged 
particles in simulated events (estimated to be also less than 0.1\%). 
The overall uncertainty is around 2\%.

\section{Calibration of the HF using \texorpdfstring{\zee}{Z to ee} events}\label{sec:hf}

The initial calibration of the HF was based on test beam data~\cite{HcalForward}.
The energy scales of the long and short fibers in six HF wedges
were set using the responses from
100\GeV electrons and negative pions.
The scale was transferred to the rest of the wedges
using radioactive source data.

The energy scale for the
long fibers is validated using events from \PZ boson decays.
The dileptonic decays of the \PZ boson are useful tools for checking the detector
calibration, because the production cross section of the \PZ bosons
at the LHC is large, and the signature is almost background free.
 The dataset used for this calibration consists of  events with
one electron candidate in the HF
and the other in the ECAL, which has been precisely calibrated.
The scale of the HF long fibers is adjusted so that the
dielectron invariant mass corresponding to the \PZ peak is consistent
between simulation and data.
The scale of the HF short fibers utilizes the short-to-long ratio measured
in an analysis of 100\GeV test beam electron data.

Because the HF is outside of the tracker acceptance,
the PF algorithm does not identify electrons that impinge on this detector
and they are instead identified as photon candidates,
which are clustered into PF jets.
An isolated high-\pt electron,
such as those produced in \PZ boson decays, would be identified as a jet
by this algorithm.
The reconstruction of HF electrons for this analysis, thus, starts with the PF jets.
To select jets consistent with being isolated electrons,
jets with characteristics consistent with
those of anomalous energy deposits are rejected.
Electron candidate jets must either have a nonzero hadronic or nonzero
electromagnetic energy after zero-suppression is applied. The electromagnetic
(EM) and hadronic (HAD) energies are defined by the deposits in the long (L)
and short (S) fibers in the following way: EM = L$-$S, \mbox{HAD = 2S}.

Jets created from energy deposits of isolated electrons have
a characteristic shower size
$\Delta R\equiv\sqrt{\smash[b]{(\Delta\eta)^2+(\Delta\phi)^2}}$ of about 0.15--0.2,
where $\Delta\phi$ is in radians.
Also, most of the energy is deposited in the long fibers,
and the energy in the short fibers is small.
Thus, HF electron candidates are required to have
a ``seed'' for the core electron shower, chosen as the
constituent of the PF jet with the largest energy.
This seed defines the initial four-vector of
the HF electron candidate. The four-vectors of other constituents of the PF jet,
ordered in energy, are added to the four-vector of the HF electron candidate if
the constituents are within $\Delta R = 0.15$ of the current four-vector of
the core shower.
The process stops when there are no more constituents close
enough to the core shower. The choice of $\Delta R = 0.15$ to sample the core
energy of the HF electron candidate is used to minimize the dependence on
{\pu}.
Because electrons from \PZ boson decays are expected to be more energetic than
electron candidates arising from the misidentification of quark
and gluon jets, the $\pt$ of HF electron candidates
are required to exceed 15\GeV.
Furthermore, the electromagnetic fraction of the candidate's energy, calculated
as a sum of all the EM energies
of the PF constituents that form
the core four-vector of the HF electron candidate, must be sufficiently large.
The EM energy is required to be nonzero, and the ratio
of the HAD
to EM energy is restricted to be less than 1.20.
The HF electron candidate is required to be isolated, as expected for leptons
from \PZ boson decays. The total energy of all the PF candidates found in the
vicinity of the shower, $0.15 < \Delta R < 0.30$, should not exceed
55\% of the energy in the core shower. These requirements result in an
efficiency of 62\% for the selection of genuine electrons from \PZ boson
decays, and retain only
1.5\% of jets misidentified as electrons, as estimated
using simulated Drell--Yan events.

Candidate \PZ boson events are selected by requiring only one
isolated electron in the ECAL with $\pt > 25$\GeV\ and with
$\abs{\eta} < 2.5$~\cite{ElectronJINST}.
The events are further required to have at least one HF electron candidate with
$\pt > 15$\GeV\ and $2.964 < \abs{\eta} < 5.191$.

\begin{figure}[htbp]
\begin{center}
  \includegraphics[width=.49\textwidth]{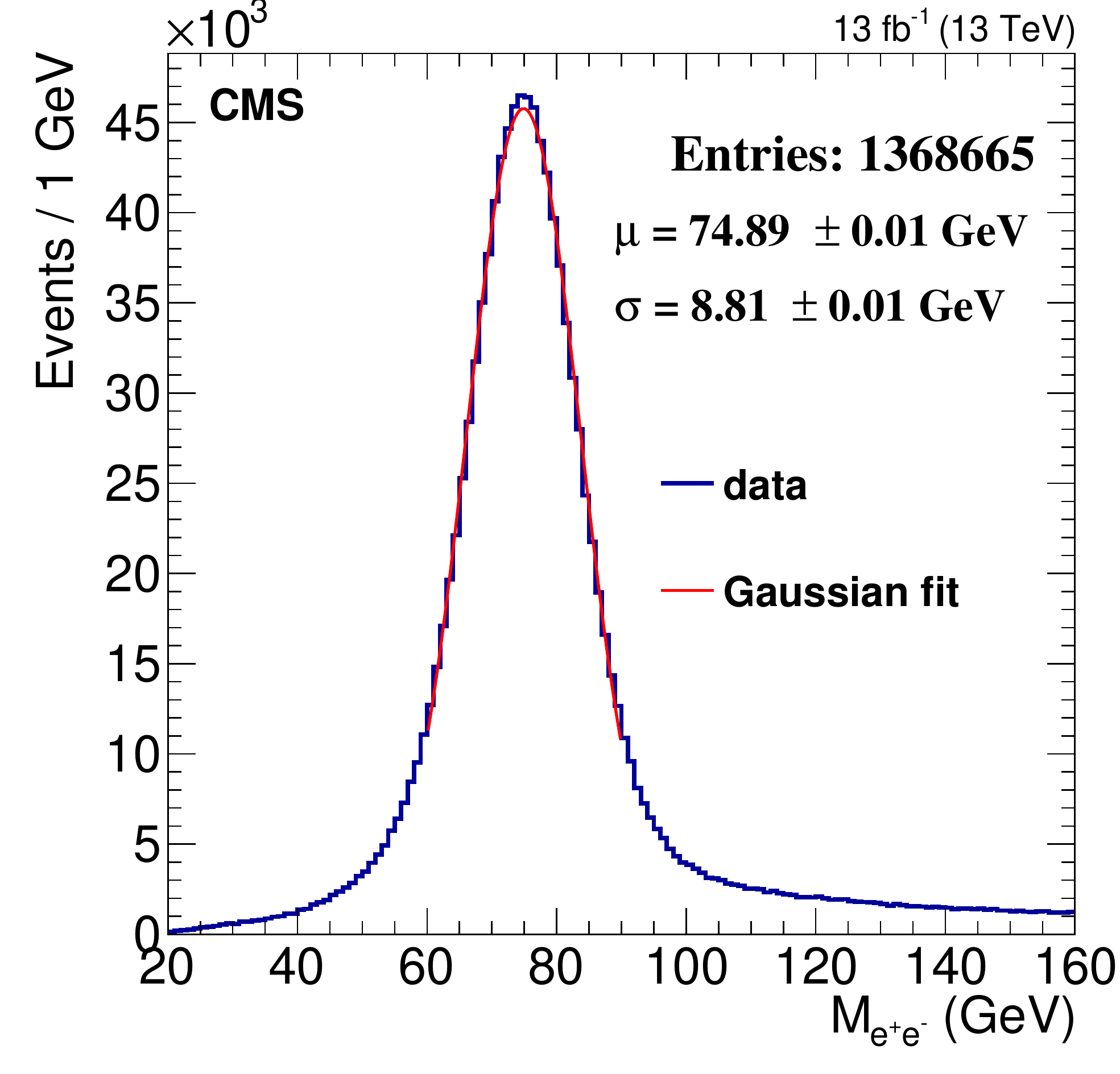}
  \includegraphics[width=.49\textwidth]{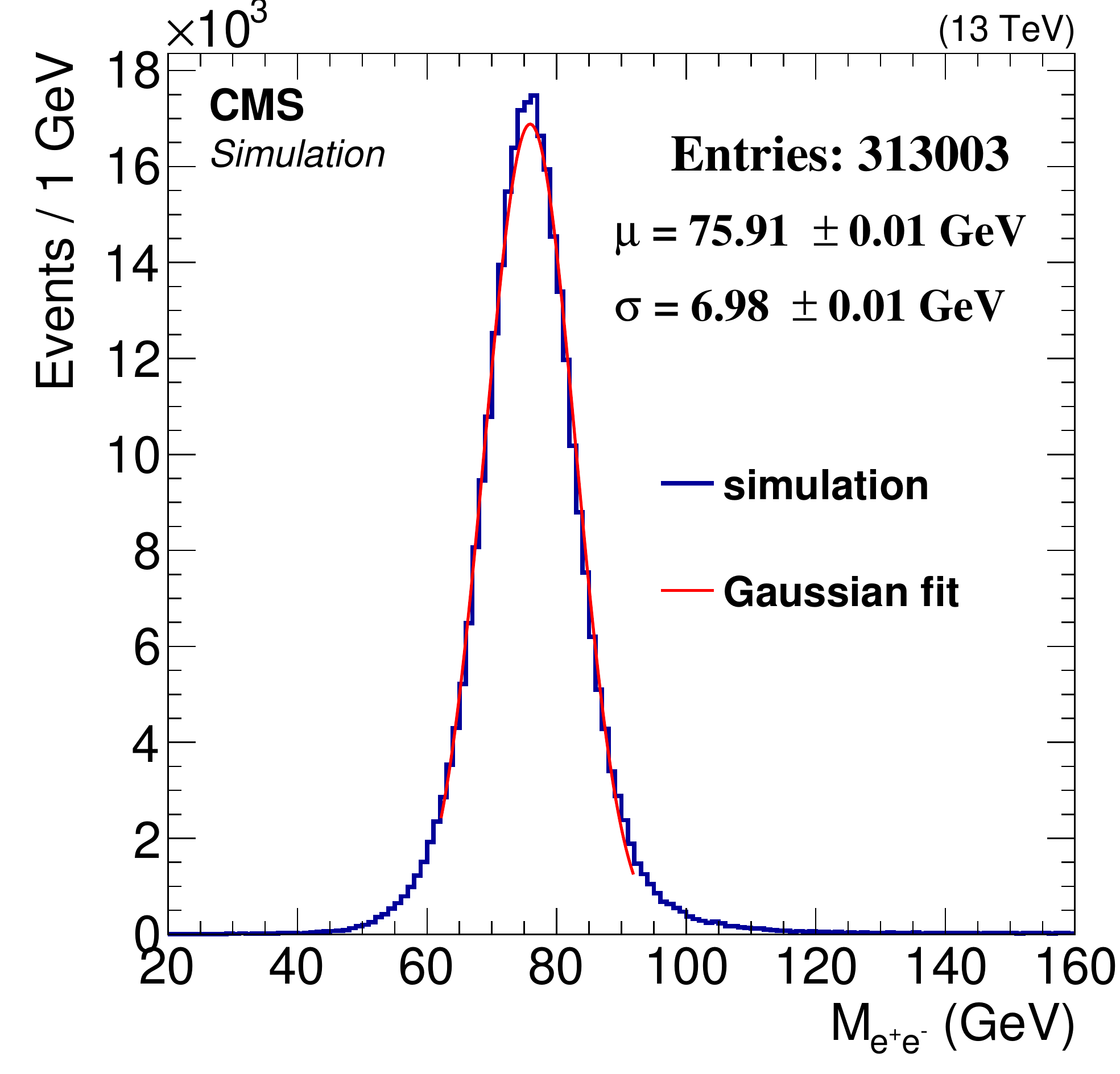}
 \end{center}
 \caption{Invariant mass of the two electrons in candidate \zee events
          for simulation (\cmsLeft), and for 2016 data (\cmsRight).
          One candidate is required to be in the ECAL, the other one in the HF.
          The mean ($\mu$) and the width ($\sigma$) of the Gaussian fits are
          shown on the plots with their uncertainties. The quality of the fits
          is sufficient to extract the mean and the width to the
          necessary accuracies.}
 \label{fig:HFMass}
\end{figure}

The dielectron invariant mass distributions ($M_{\EE}$) from candidate
\zee events are shown in Fig.~\ref{fig:HFMass} for  simulation and 2016 data.
The \PZ boson mass is measured from a fit to a Gaussian function in a restricted
mass region around the peak position.
In both simulation and data, the measured \PZ
boson mass is lower than the nominal mass (91.1876\GeV~\cite{PDG2019})
because only the long-fiber energies of towers within $\Delta R = 0.15$
of the seed  are used, resulting in an underestimation of the
shower energy due to energy leakage outside of this region.
The HF energy scale in data is not adjusted to match the fitted mean from
the invariant mass distribution using the simulation, since both values are
consistent within their uncertainties. Therefore, no corrections are applied
to the energy response in data. The 25\% difference in the width between data
and simulation is not yet
fully understood, but selection of electron pairs
in data from the \PZ boson decays, as well as shower mismodeling in the
simulation, could cause this discrepancy.

The bias in the energy due to {\pu}
is estimated using events with different numbers of {\pu} interactions.
The result is consistent with a shift of the measured
\PZ boson mass of  up to 1\%.

\begin{figure}[htbp]
\begin{center}
  \includegraphics[width=.49\textwidth]{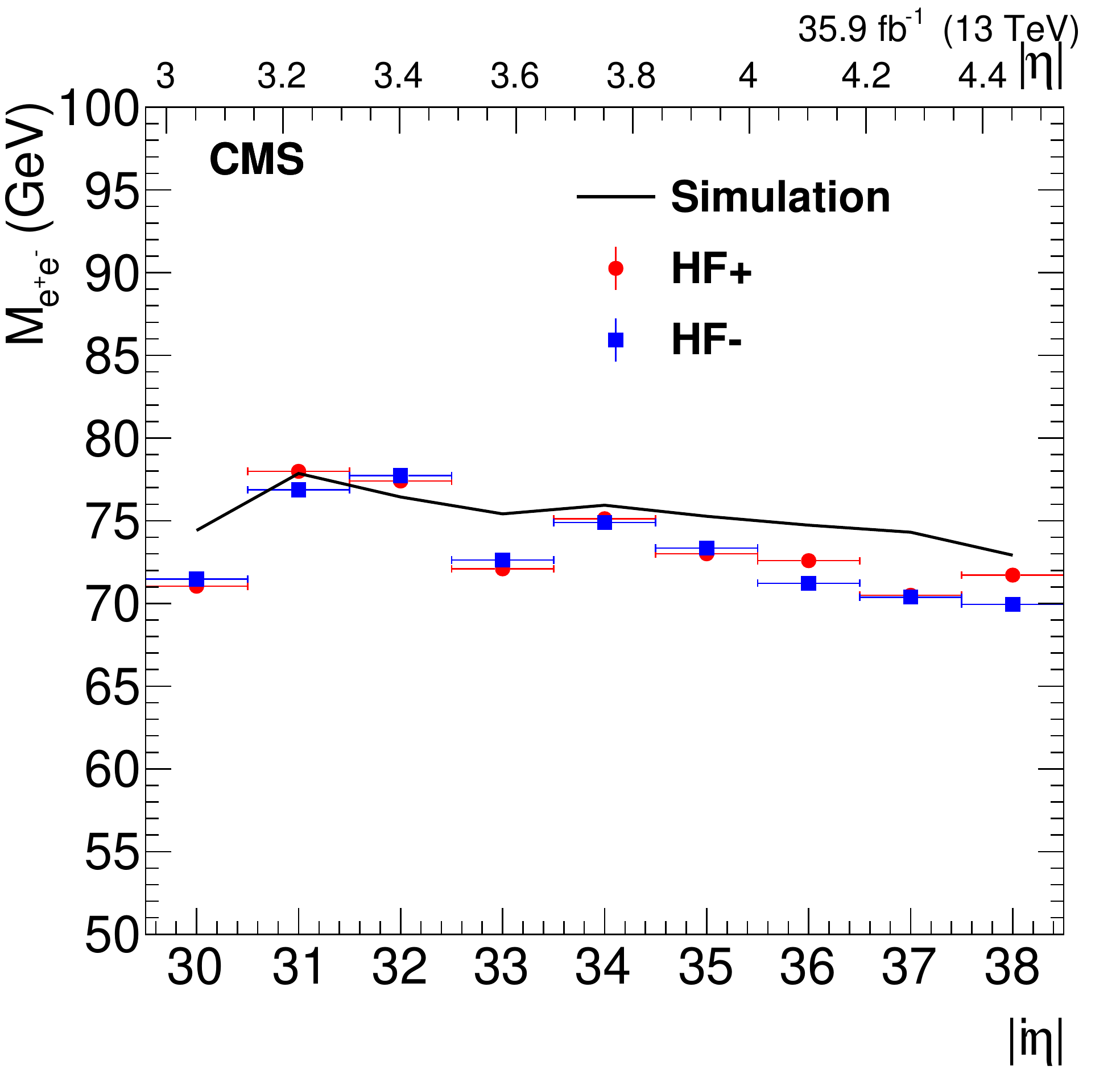}
  \includegraphics[width=.49\textwidth]{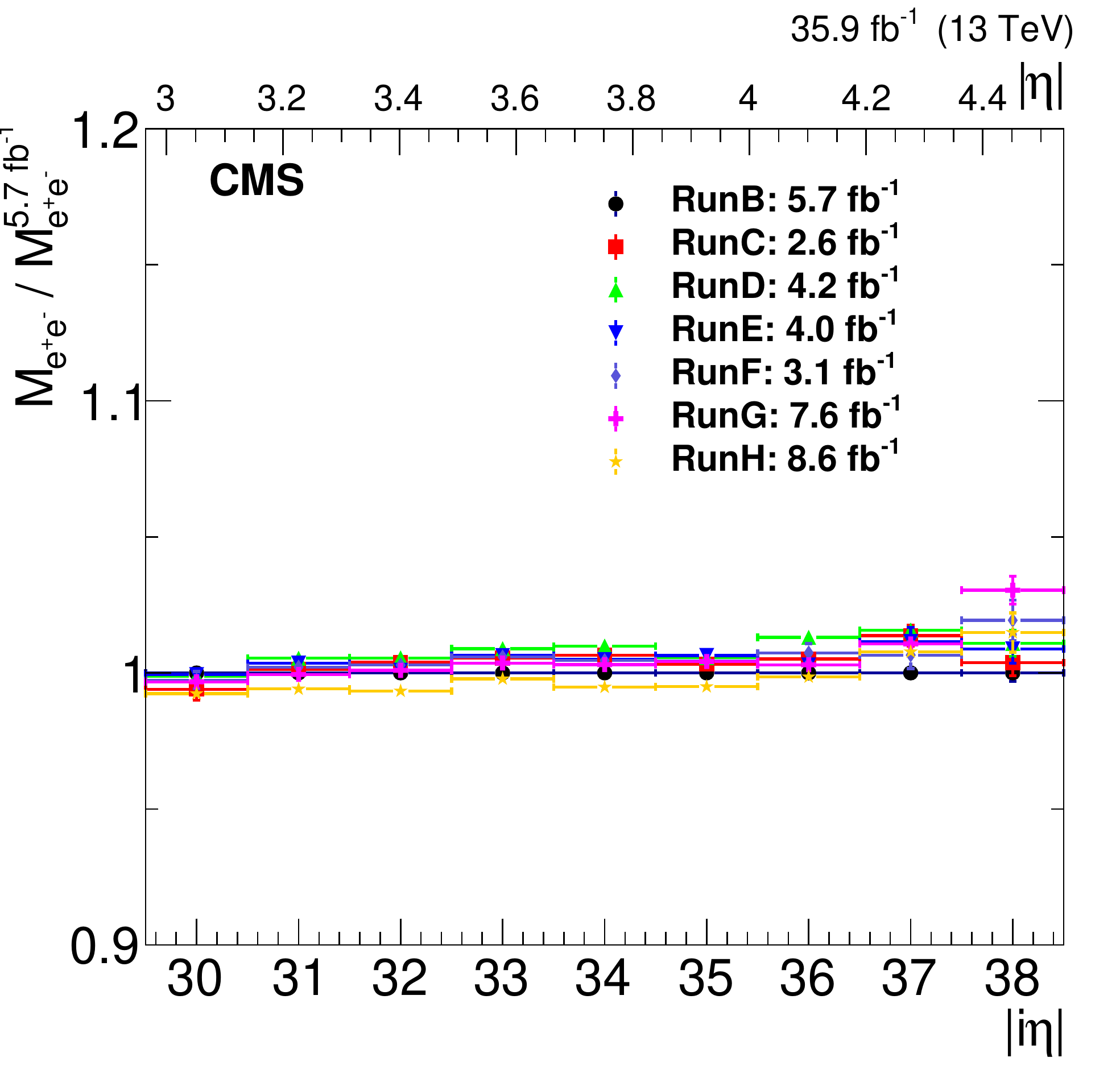}
 \end{center}
 \caption{(\cmsLeft) The results of the fits of the dielectron
          invariant mass to a Gaussian function for different
          $\eta$ values of the HF electron candidates
          obtained in simulation (black line, combined HF$+$ and HF$-$) and
          2016 data, split by HF$+$ and HF$-$ (red and blue, respectively).
          (\cmsRight) The results of the fits of the dielectron invariant
          mass to a Gaussian function for different pseudorapidity $\eta$
          values of the HF electron candidates obtained in data
          corresponding to different run ranges. The dielectron mass in
          the denominator comes from the first run range corresponding to
          5.7\fbinv. Errors on the data points are statistical only.}
 \label{fig:HFZMassb}
\end{figure}

Figure~\ref{fig:HFZMassb} shows the dielectron invariant mass as a function
of the HF electron candidate $\abs{\ieta}$, for $\ieta$ values between 30
and 38.
The HF energy response in data for both HF$+$ and HF$-$ are similar.
The energy response in the
simulation has a shape similar to that of the data, but there is a
visible trend: for the towers with $\abs{\ieta} = 33$--38, the data show
a lower energy scale than the simulation. This deviation is not completely
understood and therefore no specific corrections are considered at the moment.

\section{Calibration of the HO calorimeter}\label{sec:ho}

The calibration of the HO calorimeter is carried out in two
steps. The intercalibration makes use of muons from collision data,
as well as cosmic ray muons that traverse the tiles of the HO.
The determination of the absolute energy scale makes use of dijet events.

\subsection{Intercalibration of the HO towers}

Before data taking began,
the intercalibration of the HO towers was performed by equalizing
signals from cosmic ray muons.
This method has a few drawbacks:
\begin{itemize}
 \item large statistical uncertainties in the calibration  of the
       HO towers near $\iphi = 1$ and 37 because there are few horizontal
       muons in the cosmic muon sample;
 \item a large uncertainty in the extrapolated track position in the HO,
       which was based  on standalone muon reconstruction using information
       from muon chambers alone; this leads to a large uncertainty in the
       resulting calibration.
\end{itemize}

The intercalibration is improved using muons from \PW and \PZ decays
in $\Pp\Pp$ collisions.
Muon candidates~\cite{CMS-PAPER-MUO-10-004} are selected using the following criteria:
\begin{itemize}
\item the associated track is well measured and satisfies reconstruction quality criteria;
\item the energy in the calorimeter tower traversed by the muon is consistent with the expectation of a minimum ionizing particle;
\item the muon is isolated: no other reconstructed muon is present within a $15^{\circ}$ cone around the
muon under study and the scalar $\pt$ sum of all other tracks within $\Delta R < 0.3$ relative to the muon is less than 4\GeV;
\item the momentum of the muon after extrapolation to the HO face is more than 15, 17,
and 20\GeV\ for towers in rings 0, $\pm$1, and $\pm$2,
respectively;
\item the cosine of the angle made by the muon trajectory with respect to
the HO scintillator surface is more than 0.6 (0.4) for ring 0 (other
rings);
\item the location of the muon after extrapolation to the HO is at least 2\cm away from the
tower boundary; and
\item the time of the HO energy deposits is within 30 (20)\unit{ns}
of the beam crossing time
for towers in ring 0 (other rings).
\end{itemize}

\begin{figure}[htbp]
\begin{center}
  \includegraphics[width=.49\textwidth]{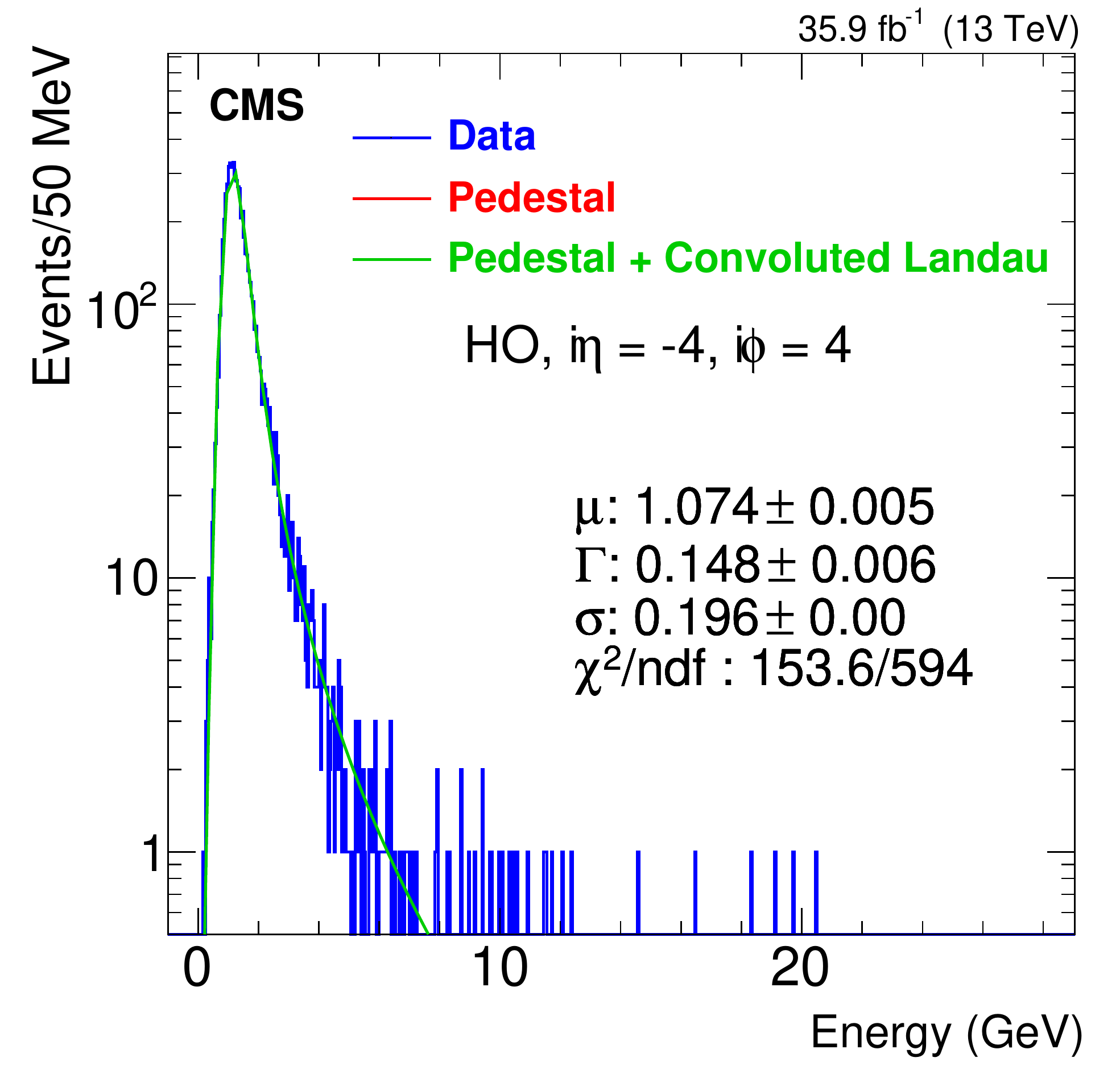}
  \includegraphics[width=.49\textwidth]{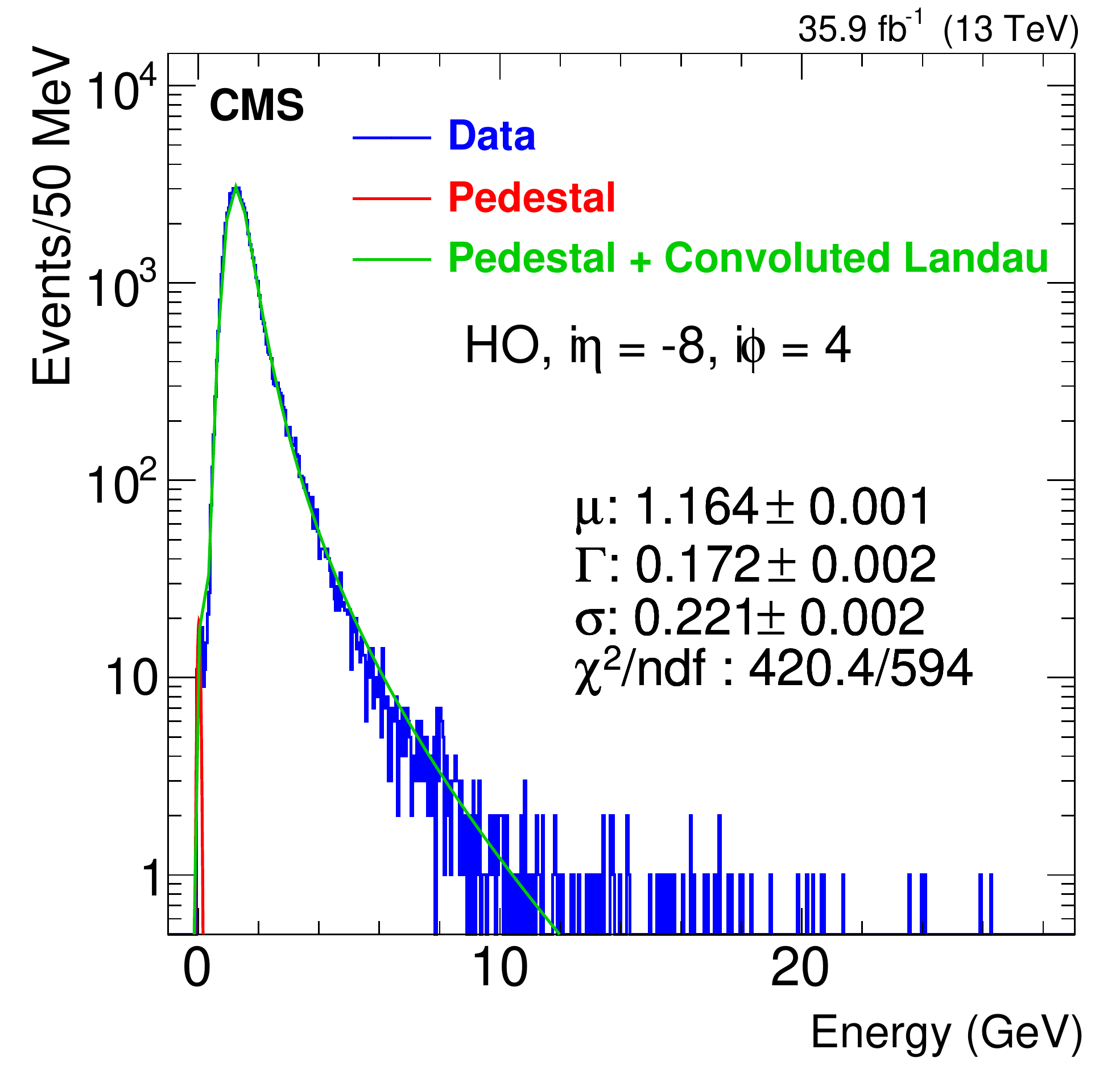}
 \end{center}
 \caption{Energy distributions for HO towers impacted by a high-\pt muon
          for the central ring ($\ieta = -4$, $\iphi = 4$, \cmsLeft)
          and for a side ring ($\ieta = -8$, $\iphi = 4$, \cmsRight),
          fitted with a combination of a Gaussian function for the pedestal
          region (shown as red lines) and a convolution of a Gaussian and a
          Landau function for the signal region (the combined fits are
          shown as the green lines).
          The parameters $\mu$, $\Gamma$ and $\sigma$ are the most probable values
          and widths of the Landau and the Gaussian functions, and ndf is
          number of degrees of freedom in the fit.}
 \label{fig:homuon1}
\end{figure}

Figure~\ref{fig:homuon1} shows a typical distribution
of the energy deposited in a tower impacted by a muon, for two typical HO towers.
The signal distribution is parametrized in the pedestal region with a Gaussian
function whose mean and width are set to values obtained from pedestal data,
and with a convolution of Gaussian and Landau functions in the signal region,
where width ($\sigma$) of the Gaussian function, the most probable value ($\mu$),
and width ($\Gamma$) of the Landau function are free parameters of the fit.
The most probable values from the fits to the convoluted function are calculated for all
towers. In parallel, the product of the path length of each muon trajectory in the
tower and the ratio of the expected energy loss of the muon with respect to the
energy loss at a fixed momentum (8\GeV) is calculated. These observed responses are
then normalized over all ($\ieta$, $\iphi$) channels. These responses are then symmetrized over all $\iphi$ channels for a given $\ieta$ ring.

\subsection{Absolute scale in the HO}

In a test beam experiment~\cite{TB06},
a 150\GeV muon beam was used to obtain the conversion factors from
charge to energy for both the HB and HO.
However, because the HO measures the energy deposited in the
tail of a hadronic shower, which typically contains low-energy
secondary particles,
the energy resolution can be improved with the application of a weight
factor to the HO energies.
This weight factor
was estimated using a test beam of 300\GeV\ $\PGpm$ mesons~\cite{HCalOuter}.

The relative weight factor derived from a single particle test beam may not be optimal for jets, so
the utility of an extra weight factor ${w_{\text{HO}}}$ is explored.
The value of ${w_{\text{HO}}}$ is tested with collision data by balancing energies in
dijet events, where one of the jets produces
a substantial energy deposit in the HO.
The analysis varies the scale of the HO energy deposits contributing to a PF jet to find the weight factor which gives the best jet energy resolution.

\begin{figure}[htbp]
\begin{center}
  \includegraphics[width=.50\textwidth]{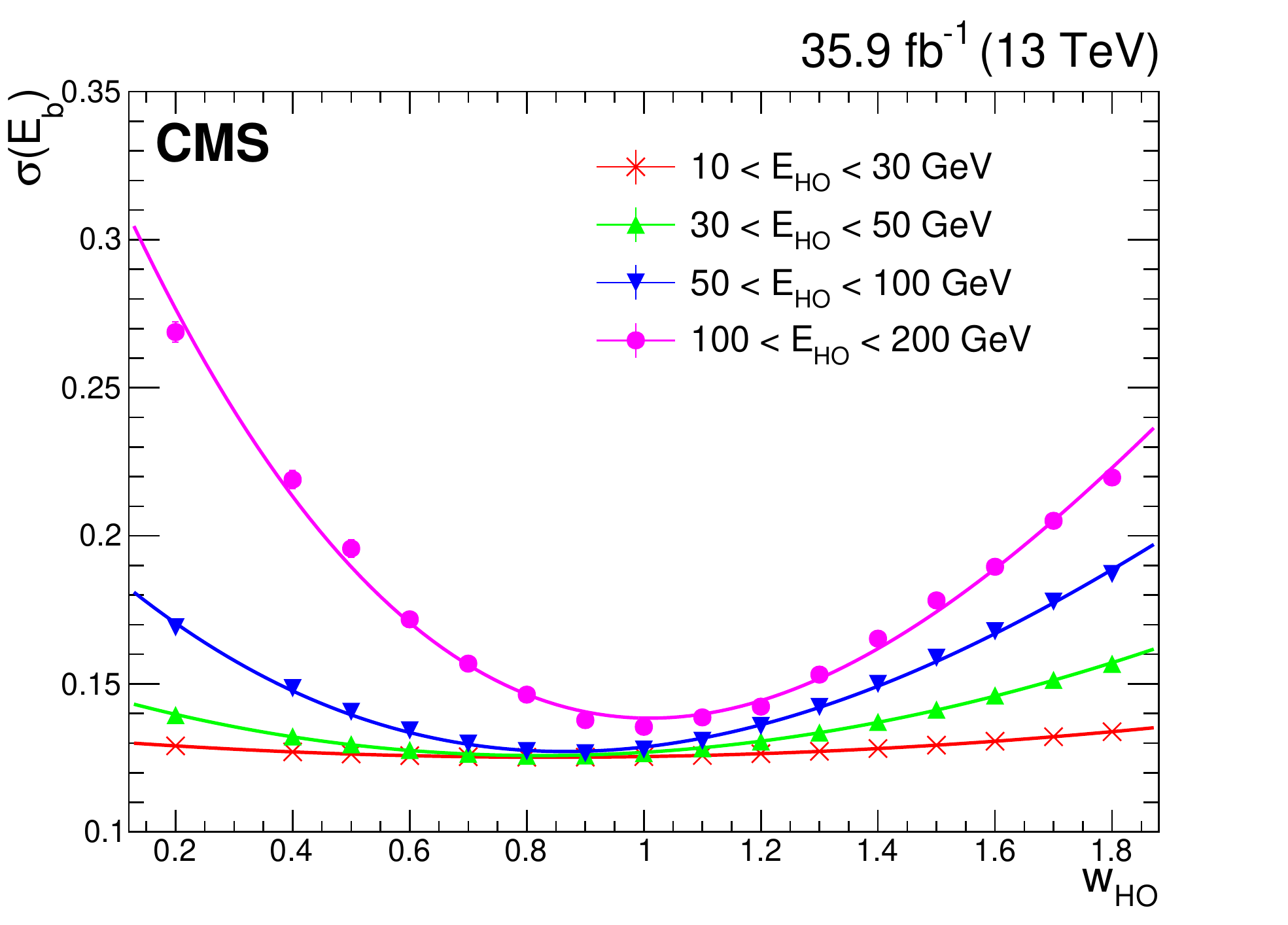}
 \end{center}
 \caption{The Gaussian width of the relative difference of $\pt$ of the two
          jets in dijet events as a function of HO weight factor in different
          ranges of energy contained in the HO cluster of the jet carrying the
          highest HO energy, when that jet is in ring 0. The smooth curves
          are results from fits to asymmetric parabolic function
          (Eq.~\eqref{eqn:sigeb}) through the points. Uncertainties are
          statistical only.}
 \label{fig:hobalance}
\end{figure}

Events are selected with the following criteria:
\begin{itemize}
 \item if there are more than two jets in the event, the third jet $\pt$ is
       required to be less than 30\GeV,
 \item the two leading jets must have $\Delta\phi > \pi /2$, and
 \item the event must not contain any isolated photon, electron, or muon with $\pt > 20$\GeV.
\end{itemize}
This analysis uses jets in $\abs{\eta} < 0.34$ (which corresponds to the central HO ring) with the highest energy HO cluster.
Figure~\ref{fig:hobalance} shows the width of the dijet energy balance distribution
as a function of ${w_{\text{HO}}}$,
where the energy balance ($E_{\mathrm{b}}$) is defined as
\begin{linenomath}
\begin{equation}
E_{\mathrm{b}}  =  2 \frac{(p_{\text{T1}} - p_{\text{T2}})}{(p_{\text{T1}} + p_{\text{T2}})},
\end{equation}
\end{linenomath}
with $p_{\text{T1}}$ and $p_{\text{T2}}$ being the transverse momenta of the leading and the
subleading jet, respectively.
The smooth curves are results from fits to asymmetric parabolic function
through the points, which is defined as,
\begin{linenomath}
\begin{equation}
\label{eqn:sigeb}
\sigma(E_{\mathrm{b}})  =  p_0 + \alpha_{1,2} \left({w_{\text{HO}}} - p_2\right)^2,
\end{equation}
\end{linenomath}
where $\alpha_{1,2}$ are used for $\left({w_{\text{HO}}} - p_2\right) \ge 0$
and $\left({w_{\text{HO}}} - p_2\right) < 0$, respectively, and $p_0$,
$\alpha_1$, $\alpha_2$, and $p_2$ are parameters of the fit.
Events containing jets with higher
HO energy provide a better sensitivity to ${w_{\text{HO}}}$.
The results imply that within the uncertainty of measurements the relative
weight factor of HO in these events is the same as in the test beam.
Similar analyses are carried out for the other HO rings ($\pm$1, $\pm$2) as well.

\section{Summary}\label{sec:summary}

The CMS experiment utilizes a variety of data to calibrate the
energy measurements obtained from its hadron calorimeter systems.
The strategy utilizes different approaches since the
calorimeter subdetectors make use of multiple technologies,
have different radiation environments, and probe a large range of particle energies.
Mean noise levels (pedestals) of all the channels are monitored for each fill of LHC and the scale factors are checked for each run period, which typically spans one month. The mean noise levels as well as the correction factors are updated whenever any of the correction factors change by more than 3\%.
The calibration is generally performed in
two steps: the azimuthal ($\phi$) intercalibration of the channels,
followed by the determination of an absolute energy scale.

In the barrel, endcap, and forward calorimeters, the $\phi$ symmetry of
minimum bias events is used to carry out an interdetector calibration, whereas
the hadron outer calorimeter utilizes reconstructed muons for this purpose.
The absolute energy calibration of the
barrel and endcap calorimeters is based on isolated charged hadrons with momenta
between 40 and 60\GeV, whereas
the calibration of the forward calorimeter relies on a validation based
on $\zee$ events.
The nonlinearity in the energy measurement of hadrons is addressed
during the particle-flow reconstruction by
using the predicted dependence on transverse momentum and pseudorapidity
from a \GEANTfour-based simulation.
Residual nonlinearities that affect the energy scale of reconstructed jets
are reduced during the calibration of the jet energy scale.
The calibration of the hadron outer calorimeter relies on the energy
balance in dijet events.

The methods use proton-proton collision data at $\sqrt{s} = 13\TeV$
collected with the CMS detector in 2016, and corresponding to integrated
luminosities up to 35.9\fbinv.
The results are applied to the final reconstruction of events
collected during that period. The systematic uncertainties in these
measurements are dominated by the uncertainties in the amount of
material between the interaction point and the detectors, including their
dependence on azimuthal angle, and by the systematic uncertainties from the
simulation of the effect of noise on the readout signal.
Systematic uncertainties in the intercalibration are below 3\% for HB, HE, HF
and HO. The uncertainty in the absolute energy scale, as determined here for
HB and HE, is below 2\%. Absolute energy scale for HF and HO are kept at
their nominal values and validated during 2016 using the methods
described in the text.
These techniques lead to the typical calibration precision of less than 3\%.

\begin{acknowledgments}
We congratulate our colleagues in the CERN accelerator departments for the excellent performance of the LHC and thank the technical and administrative staffs at CERN and at other CMS institutes for their contributions to the success of the CMS effort. In addition, we gratefully acknowledge the computing centers and personnel of the Worldwide LHC Computing Grid for delivering so effectively the computing infrastructure essential to our analyses. Finally, we acknowledge the enduring support for the construction and operation of the LHC and the CMS detector provided by the following funding agencies: BMBWF and FWF (Austria); FNRS and FWO (Belgium); CNPq, CAPES, FAPERJ, FAPERGS, and FAPESP (Brazil); MES (Bulgaria); CERN; CAS, MoST, and NSFC (China); COLCIENCIAS (Colombia); MSES and CSF (Croatia); RPF (Cyprus); SENESCYT (Ecuador); MoER, ERC IUT, PUT and ERDF (Estonia); Academy of Finland, MEC, and HIP (Finland); CEA and CNRS/IN2P3 (France); BMBF, DFG, and HGF (Germany); GSRT (Greece); NKFIA (Hungary); DAE and DST (India); IPM (Iran); SFI (Ireland); INFN (Italy); MSIP and NRF (Republic of Korea); MES (Latvia); LAS (Lithuania); MOE and UM (Malaysia); BUAP, CINVESTAV, CONACYT, LNS, SEP, and UASLP-FAI (Mexico); MOS (Montenegro); MBIE (New Zealand); PAEC (Pakistan); MSHE and NSC (Poland); FCT (Portugal); JINR (Dubna); MON, RosAtom, RAS, RFBR, and NRC KI (Russia); MESTD (Serbia); SEIDI, CPAN, PCTI, and FEDER (Spain); MOSTR (Sri Lanka); Swiss Funding Agencies (Switzerland); MST (Taipei); ThEPCenter, IPST, STAR, and NSTDA (Thailand); TUBITAK and TAEK (Turkey); NASU (Ukraine); STFC (United Kingdom); DOE and NSF (USA).

\hyphenation{Rachada-pisek} Individuals have received support from the Marie-Curie program and the European Research Council and Horizon 2020 Grant, contract Nos.\ 675440, 752730, and 765710 (European Union); the Leventis Foundation; the A.P.\ Sloan Foundation; the Alexander von Humboldt Foundation; the Belgian Federal Science Policy Office; the Fonds pour la Formation \`a la Recherche dans l'Industrie et dans l'Agriculture (FRIA-Belgium); the Agentschap voor Innovatie door Wetenschap en Technologie (IWT-Belgium); the F.R.S.-FNRS and FWO (Belgium) under the ``Excellence of Science -- EOS" -- be.h project n.\ 30820817; the Beijing Municipal Science \& Technology Commission, No. Z181100004218003; the Ministry of Education, Youth and Sports (MEYS) of the Czech Republic; the Lend\"ulet (``Momentum") Program and the J\'anos Bolyai Research Scholarship of the Hungarian Academy of Sciences, the New National Excellence Program \'UNKP, the NKFIA research grants 123842, 123959, 124845, 124850, 125105, 128713, 128786, and 129058 (Hungary); the Council of Science and Industrial Research, India; the HOMING PLUS program of the Foundation for Polish Science, cofinanced from European Union, Regional Development Fund, the Mobility Plus program of the Ministry of Science and Higher Education, the National Science Center (Poland), contracts Harmonia 2014/14/M/ST2/00428, Opus 2014/13/B/ST2/02543, 2014/15/B/ST2/03998, and 2015/19/B/ST2/02861, Sonata-bis 2012/07/E/ST2/01406; the National Priorities Research Program by Qatar National Research Fund; the Ministry of Science and Education, grant no. 3.2989.2017 (Russia); the Programa Estatal de Fomento de la Investigaci{\'o}n Cient{\'i}fica y T{\'e}cnica de Excelencia Mar\'{\i}a de Maeztu, grant MDM-2015-0509 and the Programa Severo Ochoa del Principado de Asturias; the Thalis and Aristeia programs cofinanced by EU-ESF and the Greek NSRF; the Rachadapisek Sompot Fund for Postdoctoral Fellowship, Chulalongkorn University and the Chulalongkorn Academic into Its 2nd Century Project Advancement Project (Thailand); the Nvidia Corporation; the Welch Foundation, contract C-1845; and the Weston Havens Foundation (USA).
\end{acknowledgments}

\bibliography{auto_generated}

\providecommand{\href}[2]{#2}\begingroup\raggedright\begin{thebibliography}{10}%
\makeatletter
\providecommand{\hrefCMSnoop }[0]{\@secondoftwo}%
\makeatother
\providecommand{\doi}{\texttt{doi:}\begingroup \urlstyle{tt}\Url}

\bibitem{PhysRevD93072004}
\hrefCMSnoop {}{{{CMS}} Collaboration, ``{Measurement of the top quark mass
  using proton-proton data at $\sqrt{s}$ = 7 and 8 {TeV}}'',} \textit{ Phys.
  Rev. D} \textbf{ 93} (2016) 072004,
  \href{http://dx.doi.org/10.1103/PhysRevD.93.072004}{\doi{10.1103/PhysRevD.93.072004}},
\href{http://www.arXiv.org/abs/1509.04044}{\texttt{arXiv:1509.04044}}.

\bibitem{Sirunyan2019EPJ313}
\hrefCMSnoop {}{{CMS Collaboration}, ``Measurement of the top quark mass in the
  all-jets final state at $\sqrt{s}$ = 13 {TeV} and combination with the
  lepton+jets channel'',} \textit{ Eur. Phys. J. C} \textbf{ 79} (2019) 313,
  \href{http://dx.doi.org/10.1140/epjc/s10052-019-6788-2}{\doi{10.1140/epjc/s10052-019-6788-2}},
\href{http://www.arXiv.org/abs/1812.10534}{\texttt{arXiv:1812.10534}}.

\bibitem{EPJC752352015}
\hrefCMSnoop {}{{CMS Collaboration}, ``Search for dark matter, extra
  dimensions, and unparticles in monojet events in proton-proton collisions at
  $\sqrt{s}$ = 8 {TeV}'',} \textit{ Eur. Phys. J. C} \textbf{ 75} (2015) 235,
  \href{http://dx.doi.org/10.1140/epjc/s10052-015-3451-4}{\doi{10.1140/epjc/s10052-015-3451-4}},
\href{http://www.arXiv.org/abs/1408.3583}{\texttt{arXiv:1408.3583}}.

\bibitem{1748-0221-11-10-T10005}
\hrefCMSnoop {}{{{ATLAS Tile Calorimeter System}} Collaboration, ``{The laser
  calibration of the {ATLAS} tile calorimeter during the LHC run 1}'',}
  \textit{ JINST} \textbf{ 11} (2016) T10005,
  \href{http://dx.doi.org/10.1088/1748-0221/11/10/T10005}{\doi{10.1088/1748-0221/11/10/T10005}},
\href{http://www.arXiv.org/abs/1608.02791}{\texttt{arXiv:1608.02791}}.

\bibitem{Aaboud2017}
\hrefCMSnoop {}{{ATLAS Collaboration}, ``{A measurement of the calorimeter
  response to single hadrons and determination of the jet energy scale
  uncertainty using {LHC} {Run-1} pp-collision data with the {ATLAS}
  detector}'',} \textit{ Eur. Phys. J. C} \textbf{ 77} (2017) 26,
  \href{http://dx.doi.org/10.1140/epjc/s10052-016-4580-0}{\doi{10.1140/epjc/s10052-016-4580-0}},
\href{http://www.arXiv.org/abs/1607.08842}{\texttt{arXiv:1607.08842}}.

\bibitem{TB96}
\hrefCMSnoop {}{{{CMS} {HCAL}} Collaboration, ``Studies of the response of the
  prototype {CMS} hadron calorimeter, including magnetic field effects, to
  pion, electron, and muon beams'',} \textit{ Nucl. Instrum. Meth. A} \textbf{
  457} (2001) 75,
  \href{http://dx.doi.org/10.1016/S0168-9002(00)00711-7}{\doi{10.1016/S0168-9002(00)00711-7}},
  \href{http://www.arXiv.org/abs/hep-ex/0007045}{\texttt{arXiv:hep-ex/0007045}}.

\bibitem{HCalOuter}
\hrefCMSnoop {}{{{CMS} {HCAL}} Collaboration, ``Design, performance, and
  calibration of the {CMS} hadron-outer calorimeter'',} \textit{ Eur. Phys. J.
  C} \textbf{ 57} (2008) 653,
  \href{http://dx.doi.org/10.1140/epjc/s10052-008-0756-6}{\doi{10.1140/epjc/s10052-008-0756-6}}.

\bibitem{TB06}
\hrefCMSnoop {}{{{CMS} {HCAL}/{ECAL}} Collaboration, ``The {CMS} barrel
  calorimeter response to particle beams from 2 to 350 {GeV}/$c$'',} \textit{
  Eur. Phys. J. C} \textbf{ 60} (2009) 359,
  \href{http://dx.doi.org/10.1140/epjc/s10052-009-1024-0}{\doi{10.1140/epjc/s10052-009-1024-0}}.

\bibitem{COJOCARU2004481}
\hrefCMSnoop {}{{{ATLAS Liquid Argon EMEC/HEC}} Collaboration, ``{Hadronic
  calibration of the {ATLAS} liquid argon end-cap calorimeter in the
  pseudorapidity region $1.6<|\eta|<1.8$ in beam tests}'',} \textit{ Nucl.
  Instrum. Meth. A} \textbf{ 531} (2004) 481,
  \href{http://dx.doi.org/10.1016/j.nima.2004.05.133}{\doi{10.1016/j.nima.2004.05.133}},
\href{http://www.arXiv.org/abs/physics/0407009}{\texttt{arXiv:physics/0407009}}.

\bibitem{HcalTDR}
\href {https://cds.cern.ch/record/357153/files/CMS_HCAL_TDR.pdf}{{{CMS}
  Collaboration}, ``{CMS} hadron calorimeter technical design report'',}
  Technical Report {CERN/LHCC 97-31}, 1997.

\bibitem{CMS}
\hrefCMSnoop {}{{{CMS}} Collaboration, ``The {CMS} experiment at the {CERN}
  {LHC}'',} \textit{ JINST} \textbf{ 3} (2008) S08004,
  \href{http://dx.doi.org/10.1088/1748-0221/3/08/S08004}{\doi{10.1088/1748-0221/3/08/S08004}}.

\bibitem{HcalBarrel}
\hrefCMSnoop {}{{{CMS} {HCAL}} Collaboration, ``Design, performance, and
  calibration of {CMS} hadron-barrel calorimeter wedges'',} \textit{ Eur. Phys.
  J. C} \textbf{ 55} (2008) 159,
  \href{http://dx.doi.org/10.1140/epjc/s10052-008-0573-y}{\doi{10.1140/epjc/s10052-008-0573-y}}.

\bibitem{HcalForward}
\hrefCMSnoop {}{{{CMS} {HCAL}} Collaboration, ``Design, performance and
  calibration of the {CMS} forward calorimeter wedges'',} \textit{ Eur. Phys.
  J. C} \textbf{ 53} (2008) 139,
  \href{http://dx.doi.org/10.1140/epjc/s10052-007-0459-4}{\doi{10.1140/epjc/s10052-007-0459-4}}.

\bibitem{WIRE}
\hrefCMSnoop {}{E.~Hazen {et~al.}, ``Radioactive source calibration technique
  for the {CMS} hadron calorimeter'',} \textit{ Nucl. Instrum. Meth. A}
  \textbf{ 511} (2003) 311,
  \href{http://dx.doi.org/10.1016/S0168-9002(03)01971-5}{\doi{10.1016/S0168-9002(03)01971-5}}.

\bibitem{HcalCalib1}
\hrefCMSnoop {}{{{CMS}} Collaboration, ``Performance of the {CMS} hadron
  calorimeter with cosmic ray muons and {LHC} beam data'',} \textit{ JINST}
  \textbf{ 5} (2010) T03012,
  \href{http://dx.doi.org/10.1088/1748-0221/5/03/T03012}{\doi{10.1088/1748-0221/5/03/T03012}},
  \href{http://www.arXiv.org/abs/0911.4991}{\texttt{arXiv:0911.4991}}.

\bibitem{CMS-PRF-14-001}
\hrefCMSnoop {}{{CMS Collaboration}, ``Particle-flow reconstruction and global
  event description with the {CMS} detector'',} \textit{ JINST} \textbf{ 12}
  (2017) P10003,
  \href{http://dx.doi.org/10.1088/1748-0221/12/10/P10003}{\doi{10.1088/1748-0221/12/10/P10003}},
\href{http://www.arXiv.org/abs/1706.04965}{\texttt{arXiv:1706.04965}}.

\bibitem{GEANT}
\hrefCMSnoop {}{{GEANT4} Collaboration, ``{\GEANTfour}--a simulation
  toolkit'',} \textit{ Nucl. Instrum. Meth. A} \textbf{ 506} (2003) 250,
  \href{http://dx.doi.org/10.1016/S0168-9002(03)01368-8}{\doi{10.1016/S0168-9002(03)01368-8}}.

\bibitem{JESCMS:2017}
\hrefCMSnoop {}{{CMS Collaboration}, ``{Jet energy scale and resolution in the
  {CMS} experiment in pp collisions at $8$ {TeV}}'',} \textit{ JINST} \textbf{
  12} (2017) P02014,
  \href{http://dx.doi.org/10.1088/1748-0221/12/02/P02014}{\doi{10.1088/1748-0221/12/02/P02014}},
\href{http://www.arXiv.org/abs/1607.03663}{\texttt{arXiv:1607.03663}}.

\bibitem{JINST08010}
\hrefCMSnoop {}{{{CMS}} Collaboration, ``Performance of photon reconstruction
  and identification with the {CMS} detector in proton-proton collisions at
  $\sqrt{s}$ = 8 {TeV}'',} \textit{ JINST} \textbf{ 10} (2015) P08010,
  \href{http://dx.doi.org/10.1088/1748-0221/10/08/P08010}{\doi{10.1088/1748-0221/10/08/P08010}},
  \href{http://www.arXiv.org/abs/1502.02702}{\texttt{arXiv:1502.02702}}.

\bibitem{TRK-11-001}
\hrefCMSnoop {}{{CMS Collaboration}, ``{Description and performance of track
  and primary-vertex reconstruction with the {CMS} tracker}'',} \textit{ JINST}
  \textbf{ 9} (2014) P10009,
  \href{http://dx.doi.org/10.1088/1748-0221/9/10/P10009}{\doi{10.1088/1748-0221/9/10/P10009}},
\href{http://www.arXiv.org/abs/1405.6569}{\texttt{arXiv:1405.6569}}.

\bibitem{Khachatryan:2016bia}
\hrefCMSnoop {}{{CMS Collaboration}, ``{The {CMS} trigger system}'',} \textit{
  JINST} \textbf{ 12} (2017) P01020,
  \href{http://dx.doi.org/10.1088/1748-0221/12/01/P01020}{\doi{10.1088/1748-0221/12/01/P01020}},
\href{http://www.arXiv.org/abs/1609.02366}{\texttt{arXiv:1609.02366}}.

\bibitem{Cushman}
\hrefCMSnoop {}{P.~Cushman, A.~Heering, and A.~Ronzhin, ``Custom {HPD} readout
  for the {CMS} {HCAL}'',} \textit{ Nucl. Instrum. Meth. A} \textbf{ 442}
  (2000) 289,
  \href{http://dx.doi.org/10.1016/S0168-9002(99)01236-X}{\doi{10.1016/S0168-9002(99)01236-X}}.

\bibitem{qie}
T.~M. Shaw\hrefCMSnoop {}{ {et~al.}, ``Front end readout electronics for the
  {CMS} hadron calorimeter'',} in \textit{ Nuclear Science Symposium Conference
  Record, 2002 IEEE}, p.~194.
\newblock 2002.
\newblock
  \href{http://dx.doi.org/10.1109/NSSMIC.2002.1239297}{\doi{10.1109/NSSMIC.2002.1239297}}.

\bibitem{1748-0221-11-10-T10004}
\hrefCMSnoop {}{{{CMS} {HCAL}} Collaboration, ``{Dose rate effects in the
  radiation damage of the plastic scintillators of the {CMS} hadron endcap
  calorimeter}'',} \textit{ JINST} \textbf{ 11} (2016) T10004,
  \href{http://dx.doi.org/10.1088/1748-0221/11/10/T10004}{\doi{10.1088/1748-0221/11/10/T10004}},
\href{http://www.arXiv.org/abs/1608.07267}{\texttt{arXiv:1608.07267}}.

\bibitem{Dugad}
R.~A. Shukla\hrefCMSnoop {}{ {et~al.}, ``{Microscopic characterisation of photo
  detectors from {CMS} hadron calorimeter}'',} \textit{ Rev. Sci. Instrum.}
  \textbf{ 90} (2019) 023303,
  \href{http://dx.doi.org/10.1063/1.5046465}{\doi{10.1063/1.5046465}},
\href{http://www.arXiv.org/abs/1806.09887}{\texttt{arXiv:1806.09887}}.

\bibitem{Collaboration_2010}
\hrefCMSnoop {}{{CMS Collaboration}, ``Identification and filtering of
  uncharacteristic noise in the {CMS} hadron calorimeter'',} \textit{ JINST}
  \textbf{ 5} (2010) T03014,
  \href{http://dx.doi.org/10.1088/1748-0221/5/03/t03014}{\doi{10.1088/1748-0221/5/03/t03014}}.

\bibitem{Sirunyan_2019}
\hrefCMSnoop {}{{CMS Collaboration}, ``Performance of missing transverse
  momentum reconstruction in proton-proton collisions at $\sqrt{s}$ = 13 {TeV}
  using the {CMS} detector'',} \textit{ JINST} \textbf{ 14} (2019) P07004,
  \href{http://dx.doi.org/10.1088/1748-0221/14/07/p07004}{\doi{10.1088/1748-0221/14/07/p07004}}.

\bibitem{Cacciari:2008gp}
\hrefCMSnoop {}{M.~Cacciari, G.~P. Salam, and G.~Soyez, ``The anti-\kt jet
  clustering algorithm'',} \textit{ JHEP} \textbf{ 04} (2008) 063,
  \href{http://dx.doi.org/10.1088/1126-6708/2008/04/063}{\doi{10.1088/1126-6708/2008/04/063}},
  \href{http://www.arXiv.org/abs/0802.1189}{\texttt{arXiv:0802.1189}}.

\bibitem{Cacciari:2011ma}
\hrefCMSnoop {}{M.~Cacciari, G.~P. Salam, and G.~Soyez, ``{FastJet user
  manual}'',} \textit{ Eur. Phys. J. C} \textbf{ 72} (2012) 1896,
  \href{http://dx.doi.org/10.1140/epjc/s10052-012-1896-2}{\doi{10.1140/epjc/s10052-012-1896-2}},
\href{http://www.arXiv.org/abs/1111.6097}{\texttt{arXiv:1111.6097}}.

\bibitem{Sjostrand:2014zea}
T.~Sj{\"o}strand\hrefCMSnoop {}{ {et~al.}, ``{An introduction to Pythia
  8.2}'',} \textit{ Comput. Phys. Commun.} \textbf{ 191} (2015) 159,
  \href{http://dx.doi.org/10.1016/j.cpc.2015.01.024}{\doi{10.1016/j.cpc.2015.01.024}},
\href{http://www.arXiv.org/abs/1410.3012}{\texttt{arXiv:1410.3012}}.

\bibitem{Alwall:2014hca}
J.~Alwall\hrefCMSnoop {}{ {et~al.}, ``{The automated computation of tree-level
  and next-to-leading order differential cross sections, and their matching to
  parton shower simulations}'',} \textit{ JHEP} \textbf{ 07} (2014) 079,
  \href{http://dx.doi.org/10.1007/JHEP07(2014)079}{\doi{10.1007/JHEP07(2014)079}},
\href{http://www.arXiv.org/abs/1405.0301}{\texttt{arXiv:1405.0301}}.

\bibitem{Ball:2014uwa}
\hrefCMSnoop {}{{NNPDF} Collaboration, ``Parton distributions for the {LHC Run
  II}'',} \textit{ JHEP} \textbf{ 04} (2015) 040,
  \href{http://dx.doi.org/10.1007/JHEP04(2015)040}{\doi{10.1007/JHEP04(2015)040}},
\href{http://www.arXiv.org/abs/1410.8849}{\texttt{arXiv:1410.8849}}.

\bibitem{Khachatryan:2015pea}
\hrefCMSnoop {}{{CMS Collaboration}, ``{Event generator tunes obtained from
  underlying event and multiparton scattering measurements}'',} \textit{ Eur.
  Phys. J. C} \textbf{ 76} (2016) 155,
  \href{http://dx.doi.org/10.1140/epjc/s10052-016-3988-x}{\doi{10.1140/epjc/s10052-016-3988-x}},
\href{http://www.arXiv.org/abs/1512.00815}{\texttt{arXiv:1512.00815}}.

\bibitem{ElectronJINST}
\hrefCMSnoop {}{{CMS Collaboration}, ``{Performance of electron reconstruction
  and selection with the {CMS} detector in proton-proton collisions at
  $\sqrt{s}= 8$ {TeV}}'',} \textit{ JINST} \textbf{ 10} (2015) P06005,
  \href{http://dx.doi.org/10.1088/1748-0221/10/06/P06005}{\doi{10.1088/1748-0221/10/06/P06005}},
\href{http://www.arXiv.org/abs/1502.02701}{\texttt{arXiv:1502.02701}}.

\bibitem{PDG2019}
\hrefCMSnoop {}{{Particle Data Group}, ``{Review of particle physics}'',}
  \textit{ Phys. Rev. D} \textbf{ 98} (2018) 030001,
  \href{http://dx.doi.org/10.1103/PhysRevD.98.030001}{\doi{10.1103/PhysRevD.98.030001}}.

\bibitem{CMS-PAPER-MUO-10-004}
\hrefCMSnoop {}{{{CMS}} Collaboration, ``Performance of {CMS} muon
  reconstruction in pp collision events at {$\sqrt{s} = 7${TeV}}'',} \textit{
  JINST} \textbf{ 7} (2012) P10002,
  \href{http://dx.doi.org/10.1088/1748-0221/7/10/P10002}{\doi{10.1088/1748-0221/7/10/P10002}},
  \href{http://www.arXiv.org/abs/1206.4071}{\texttt{arXiv:1206.4071}}.

\end{thebibliography}\endgroup
\cleardoublepage \appendix\section{The CMS Collaboration \label{app:collab}}\begin{sloppypar}\hyphenpenalty=5000\widowpenalty=500\clubpenalty=5000\vskip\cmsinstskip
\textbf{Yerevan Physics Institute, Yerevan, Armenia}\\*[0pt]
A.M.~Sirunyan$^{\textrm{\dag}}$, A.~Tumasyan
\vskip\cmsinstskip
\textbf{Institut f\"{u}r Hochenergiephysik, Wien, Austria}\\*[0pt]
W.~Adam, F.~Ambrogi, T.~Bergauer, J.~Brandstetter, M.~Dragicevic, J.~Er\"{o}, A.~Escalante~Del~Valle, M.~Flechl, R.~Fr\"{u}hwirth\cmsAuthorMark{1}, M.~Jeitler\cmsAuthorMark{1}, N.~Krammer, I.~Kr\"{a}tschmer, D.~Liko, T.~Madlener, I.~Mikulec, N.~Rad, J.~Schieck\cmsAuthorMark{1}, R.~Sch\"{o}fbeck, M.~Spanring, D.~Spitzbart, W.~Waltenberger, C.-E.~Wulz\cmsAuthorMark{1}, M.~Zarucki
% ...
\vskip\cmsinstskip
\dag: Deceased\\
1:  Also at Vienna University of Technology, Vienna, Austria\\
% ...
\end{sloppypar}
\end{document}